	\documentclass[12pt, draftclsnofoot, onecolumn]{IEEEtran}

	\usepackage{pifont,bm,multicol,amsfonts,amsmath,color,amssymb,graphicx, epsfig,cite,psfrag,subfigure,algorithm,enumerate,stfloats,algorithm,algorithmic,epstopdf,balance}
\usepackage{tabularx}
	\usepackage{pifont,bm,multicol,amsfonts,amsmath,color,amssymb,graphicx, epsfig,cite,psfrag,subfigure,algorithm,enumerate,stfloats,algorithm,algorithmic,epstopdf,balance}

	\usepackage{setspace}

	\newtheorem{remark}{\underline{Remark}}

	\def\tr{\mathop\mathrm{tr}}

\begin{document}
\title{Time-Varying Downlink Channel Tracking for   Quantized Massive MIMO Networks} 	
\author{Jianpeng Ma,  Shun Zhang, { \emph{Member, IEEE,}} Hongyan Li, \emph{Senior Member, IEEE,}  \\ Feifei Gao,{ \emph{Senior Member, IEEE}}, and Zhu Han,{ \emph{Fellow, IEEE}}
	
    \thanks{J. Ma, S. Zhang, H. Li are with the State Key Laboratory of Integrated Services Networks, Xidian University, Xi¡¯an 710071, P. R. China (Email: jpma@xidian.edu.cn; zhangshunsdu@xidian.edu.cn;
hyli@xidian.edu.cn).}

    \thanks{F. Gao is with the Tsinghua National Laboratory for Information Science and Technology (TNList), Tsinghua University, Beijing 100084, China (Email: feifeigao@ieee.org).}
    \thanks{H. Zhu is with the Electrical and Computer Engineering Department, University of Houston, Houston, TX, 77004 USA (Email:
zhan2@uh.edu).}
}
\maketitle
  \begin{abstract}

    This paper proposes a Bayesian downlink channel estimation algorithm for   time-varying     massive MIMO networks. In particular, the quantization  effects  at the receiver are considered.
   In order to \textcolor{black}{fully exploit the sparsity and time correlations of   channels,}
   we formulate  the time-varying massive MIMO channel as \textcolor{black}{the} simultaneously sparse signal model. Then, we propose a  sparse
   Bayesian learning (SBL) framework to learn  the model parameters of the sparse virtual channel.   To reduce complexity, we employ    the expectation maximization (EM) algorithm   to achieve the  approximated  solution.  Specifically, the factor graph and the general approximate message passing (GAMP) algorithms are used to compute the  desired posterior statistics in the expectation step, so that   high-dimensional integrals over the marginal distributions can be avoided.
   The non-zero supporting vector of a virtual channel is then obtained from channel statistics by a k-means clustering algorithm.   After that, the  reduced  {dimensional} \textcolor{black}{GAMP-based scheme} is applied  to make the full use of  the channel temporal correlation so as to   enhance the virtual channel tracking  {accuracy}.
  Finally, we demonstrate the	efficacy of the proposed schemes through simulations.
\end{abstract}
\newpage
\section{Introduction}

Using a large number of antennas at the base station (BS), 	the massive multiple-input multiple-output (MIMO)  has   outstanding advantages  in spectral efficiency and power efficiency \cite{efficiency2,Massive_in_5G_1,Massive_in_5G_2}. Since
both the  downlink precoding and the uplink detection  need the  accurate channel state information (CSI), the   performance of massive MIMO heavily relies on  the CSI at the BS.
The CSI   can be obtained through  the  uplink training in  the  time-division duplex (TDD) systems, where the  uplink-downlink reciprocity exists\cite{TDD-massive,feedback}.
In  the frequency-division duplex  (FDD) system,  the CSI should be obtained through downlink training, user estimation, and feedback.
{Correspondingly,
	the overhead  of training is  in scale with the number of   antennas at the BS, so is the CSI  feedback overhead \cite{training_fdd,training_fdd2,Zhang2014Power}. However, due to the advantage of the FDD mode for the long multipath scenarios, the FDD mode still plays an important role in the present  cellular systems\cite{JSDM}.

The precoding  and signal detection  of massive MIMO in the FDD mode have been well studied.  Then, reducing the overhead of channel acquisition has become the recently hot topic \cite{JSDM,JSDM_Opportunistic,JSDM_mm,beam_division,prebeamform2,jpmatwc,Gao2015,gao2015spatially,VCR,gao_jsac}. One  common approach is to fully exploit the sparsity of the massive MIMO channel to reduce
the number of the effective channel parameters.
 From various measurement campaigns about massive MIMO channels at the millimeterwave band,
 we can find that the scattering effect of the environment is limited in one narrow angle spread region\cite{JSDM}.
 Thus, the wireless channel can be sparsely reformulated in the angular domain.  Many previous works have proposed efficient downlink channel estimation and feedback algorithms based on this sparse assumption.
Generally speaking, there are three main methods in literature:

{\it 1) Singular value decomposition (SVD)}\text{\cite{JSDM,JSDM_Opportunistic,JSDM_mm,beam_division,prebeamform2,jpmatwc}}: SVD based methods exploits the low-rank property of  {the} massive MIMO  channel covariance matrix.
However,   SVD for the high-dimensional  covariance matrix has high computational complexity. Moreover, the acquisition of channel covariance matrix is not easy.

{\it 2) Compressive sensing (CS)}\cite{Gao2015,gao2015spatially}: When the channel can be sparsely represented, CS-based techniques can robustly recover the sparse signal with reduced overhead. However, the computational complexity of CS based methods is still high.

{\it 3) Virtual channel representation (VCR)}\cite{VCR,jpmatcom,gao_jsac}: When  the BS is equipped with a massive uniform linear or rectangular, the  discrete Fourier transform (DFT) of the channel vector (called as virtual channel) contains many zero elements. The main task of this method is to obtain the positions of the non-zero elements within the virtual channel.

Although the above mentioned works  can  effectively  reduce the overhead of   channel training and CSI feedback, they only consider the static or quasi-static fading massive MIMO   channels.   To the best of our knowledge, there are just a very limited number of studies  on time-varying massive MIMO channel estimation. In \cite{time-vary-gao1},   the authors proposed  a spatial and temporal basis
expansion model (BEM) to reduce the effective dimensions of the channels, where the spatial channel is decomposed into the time-varying spatial information and the time-varying gain information.
In \cite{jpmatcom}, the authors proposed a  channel estimation scheme for the time-varying TDD massive MIMO networks, where the Kalman filter (KF) and  the  Rauch-Tung-Striebel smoother (RTSS) are  {utilized} to track   the posterior statistics of the sparse channel.
In \cite{iterative},  the authors developed a  {low-complex} online iterative algorithm to track the beamformer for massive MIMO systems.  A compensation technique to offset the variation of the time-varying optimal solution was proposed.
 In \cite{Guvensen}, the authors  assumed that the  channel was a   stationary Gauss-Markov random process, and a reduced rank   Kalman filtering based prebeamformer design  method is  proposed for the TDD systems.

   In this paper, we  propose a Bayesian downlink channel estimation algorithm for     time-varying     massive MIMO networks in the FDD mode. In particular, the quantization effects  at the receiver are considered.
   In order to fully exploit the channel sparsity to reduce the overhead and utilize the channel temporal correlation to enhance the estimation accuracy, we formulate  the time-varying massive MIMO channel as  a simultaneously sparse signal model with
    the help of both the virtual channel representation (VCR) and  the first order auto regressive (AR) model. Then, we propose a  sparse
   Bayesian learning (SBL) framework \cite{kwong} to learn  the model parameters of the sparse virtual channel.   To reduce complexity, we apply    the expectation maximization (EM) algorithm   to achieve the  approximate  solution.  Specifically, the factor graph and the message passing algorithms are used to compute the desired posterior statistics in the expectation step, so that   high-dimensional integrals over the marginal distributions can be avoided.
   The non-zero supporting vector of the virtual channel is then obtained from channel statistics by a k-means clustering algorithm.
   After parameter learning, we construct the dynamical state-space model for the virtual channel tracking,
  and
    design the  reduced  {dimensional} \textcolor{black}{GAMP-based scheme} to make the full use of  the channel temporal correlation and   enhance the virtual channel tracking  {accuracy}.

The rest of this paper is organized as follows.
Section II introduces the system configuration  and   time-varying sparse virtual channel model, and presents a summary of quantization.
 In Section III  we investigate how to learn the model parameters of the sparse virtual channel.
  The virtual channel tracking is presented in Section IV.    The  Simulation results are presented  in Section   V, and the conclusions are drawn in Section VI.

Notations:  We  use lowercase (uppercase)  boldface to  denote  vector (matrix).
$(\cdot)^T$, $(\cdot)^*$, and $(\cdot)^H $ represent  the  transpose,  the
complex conjugate and  the  Hermitian transpose, respectively. $\mathbf{I}_N$
representes a $N\times N$ identity matrix. $\delta(\cdot)$ is the Dirac delta function.  $\mathbb E \{\cdot\}$  {is} the  expectation operator. We use $\text{tr}\{\cdot\}$, $\det\{\cdot\}$ and $\text{rank}\{\cdot\}$  to  denote the  trace,  the  determinant, and  the  rank of a matrix, respectively. $[\mathbf{X}]_{ij}$ is the $(i,j)$-th entry  of  $\mathbf{X}$. $\mathbf X_{:,\mathcal Q}$  ( or $\mathbf X_{\mathcal Q,:}$) is the submatrix of $\mathbf{X}$  {and contains} the columns (or rows) with  {the} index set $\mathcal Q$.  $\mathbf x_{\mathcal Q}$ is the subvector of $\mathbf{x}$  {formed by} the entries   with  {the} index set $\mathcal Q$.  $\mathbf{n} \sim \mathcal{CN}(0,\mathbf{I}_{N})$ means  that  $\mathbf{n}$ is complex circularly-symmetric Gaussian distributed with zero mean and covariance $\mathbf{I}_{N}$. $\lfloor x \rfloor$ denotes the smallest integer no less than $x$,  {while} $\lceil x \rceil$  {represents} the largest integer no more than $x$. $\backslash$ is the set subtraction operation. $\Re(x)$  {is} the real component  of $x$. $\text{diag}(\mathbf X)$ is a column vector formed  {by}
the diagonal  {elements} of $\mathbf X$.

\section{System Model}

In this work, we will consider a single-cell massive MIMO system, where the
BS is equipped with $ N\gg1 $ antennas \textcolor{black}{in the form of the} the uniform linear array (ULA).
\textcolor{black}{$K$ users with} single-antenna \textcolor{black}{are} randomly distributed in the coverage area.
\textcolor{black}{We assume that  the channel are quasi-static during  a block of $L$ channel uses  and changes from block to block.}
 Similar to \cite{Fleury,Raghavan}
we will utilize the physical channel model to describe the inherent sparsity
and the temporal correlation for the massive MIMO channels. Then, \textcolor{black}{during
$m$-th time block}, the physical DL channel  from the BS to the user $k$  can be written as
\begin{align}
\mathbf h_{k,m}=\int_{-\infty}^{+\infty}\int_{\theta_k^{\text{min}}}^{\theta_k^{\text{max}}}
\mathbf a(\theta) e^{\jmath 2\pi \nu mL T_s}
\hbar_{k}(\theta,\nu) d\theta d\nu,\label{eq:h_k}
\end{align}
where $\hbar_k(\theta,\nu)$ is the joint
angle-Doppler channel gain function of   user $k$ corresponding to
the  direction of departure (DOD) $\theta$ and   Doppler frequency $\nu$,
and $\frac{1}{T_s}$ is the system sampling rate.
\textcolor{black}{Moreover, $\mathbf a(\theta)$ denotes the BS's  array response vector with respect
to the emergence angle $\theta$ and can be defined as}
\begin{align}\label{steer_vec}
\mathbf a(\theta)=\Big[1,e^{\jmath 2\pi \frac{d}{\lambda}\sin(\theta)},\ldots,e^{\jmath2\pi(N-1)\frac{ d}{\lambda}\sin(\theta)}\Big]^T,
\end{align}
\textcolor{black}{where $\lambda$  is the  signal  carrier wavelength, and $d$ represents the antenna spacing.}
The channels from the BS to different users are assumed to be statistically independent.

As in  \cite{JIN_MMWAVE}, the  VCR
can be utilized to dig the the sparsity of $\mathbf h_{k,m}$ as}
\begin{equation}
\mathbf{\widetilde {h}}_{k,m}=\mathbf {F}_{N}\mathbf {h}_{k,m},\label{eq:virtual_channel}
\end{equation}
 where $\mathbf{\widetilde {h}}_{k,m}$ is the virtual channel of $\mathbf {h}_{k,m}$, and $\mathbf F_{N}$ is the $N\times N$ normalized DFT matrix with the $(i,j)$th entry as $[\mathbf {F}_{N}]_{i,j}=\frac {1}{\sqrt N}e^{-j\frac{2\pi ij}{N}}$.
 \textcolor{black}{It can be checked from (\ref{eq:virtual_channel})
that the locations of the non-zero elements of $\mathbf {h}_{k,m}$ depends on the angle spread (AS) information
of the user $k$, i.e., $[\theta_k^{\min}, \theta_k^{\max}]$.
Theoretically, the AS information does not change drastically within
thousands of the channel coherence time $LT_s$, which means that the non-zero supporting vector for $\mathbf{\widetilde {h}}_{k,m}$ will remain time-invariant
within a much longer period. Furthermore, under the massive MIMO scenario, especially at
the millimeterwave and Tera Hertz bands, the AS will be limited in one narrow region,
and the number of the non-zero elements in $\mathbf{\widetilde {h}}_{k,m}$, will be much less than $N$.
Consequently, the virtual channel $\mathbf{\widetilde h}_{k,m}$ can be treated as suitably sparse signal.}

To capture the sparsity of $\mathbf{\widetilde h}_{k,m}$, we can adopt
the Gaussian scale mixture function to describe the prior PDF
$p(\mathbf{\widetilde h}_{k,m})$ as
\begin{align}
p(\mathbf{\widetilde h}_{k,m})=\prod\limits_{i=1}^N \mathcal{CN}([\mathbf{\widetilde h}_{k,m}]_i;0,\lambda_{k,i})p(\lambda_{k,i}), \label{eq:prior_h}
\end{align}
where the hyperprior $p(\lambda_{k,i})$ represents the mixing density and controls the
sparsity of $p(\mathbf{\widetilde h}_{k,m})$. Without loss of generality, the exponential density will be utilized
for $p(\lambda_{k,i})$. \textcolor{black}{Furthermore, we will utilize the first order auto regressive model
 to characterize the time-correlation of $\mathbf{\widetilde {h}}_{k,m}$ as}
 \begin{align}
 \mathbf{\tilde  h}_{k,m}=&\alpha_k \mathbf{ \tilde h}_{k,m-1}+ \sqrt{1-\alpha_k^2}\boldsymbol\upsilon_{k,m},
 \label{eq:state}
 \end{align}
where $\alpha_{k}$ is the transmission factor and depicts the time-correlation property, $\boldsymbol\upsilon_{k,m}\sim\mathcal{CN}(0,\boldsymbol \Lambda_{k})$ is the noise vector, the $N\times N$ diagonal matrix $\boldsymbol \Lambda_{k} = \text{diag}(\underbrace{[\lambda_{k,1},\lambda_{k,2},\cdots, \lambda_{k,N}]^T}_{\boldsymbol\lambda_k})$.

Furthermore, we   consider the effects of the quantization at the receiver\cite{Fan2015Uplink,fangjun,wang2017bayesian}. Specially,
the discrete quantization function of the complex value $x$, i.e., $\mathcal Q(x)$, can be written as
\begin{align}
\mathcal{Q}(x)=k_1+jk_2,~\text{for}~\epsilon_{k_1}^{L}\le\Re\{{x}\}<\epsilon_{k_1}^{U},~
\epsilon_{k_2}^{L}\le\Im\{{x}\}<\epsilon_{k_2}^{U},
\end{align}
where   integer numbers $k_1$ and $ k_2$ lie  within the integer set
$\left\{-\frac{2^{\kappa}}{2}+1, -\frac{2^{\kappa}}{2}+2,\ldots, \frac{2^{\kappa}}{2}\right\}$,
and $\kappa$ represents the number of the quantization bits.
$\epsilon_{k_1}^{L}$ and $\epsilon_{k_1}^{U}$ are separately the low and up detection threshold with
respect to the discrete out $k_1$, and can be defined as
\begin{align}
\epsilon_{k_1}^{L}=\left\{\begin{aligned}
&\left(k_1-\frac{1}{2}\right)\Delta, &~k_1\ge -\frac{2^{\kappa}}{2},\\
&-\infty, &~\text{otherwise},
\end{aligned}
\right.
\kern 30pt
\epsilon_{k_1}^{U}=\left\{\begin{aligned}
&\left(k_1+\frac{1}{2}\right)\Delta, &~k_1\le \frac{2^{\kappa}}{2}-1,\\
&+\infty, &~\text{otherwise},
\end{aligned}
\right.\label{eq:nor_quanti}
\end{align}
and $\Delta$ represents the fixed quantization step size.

For the pseudo-de-Quantization (PDQ),
$\mathcal Q(x)$ can be reexpressed as
\begin{align}
\mathcal Q(x)=(1-\rho)x+n_q,\label{eq:pseu_quanti}
\end{align}
where $\rho$ is distortion factor, and $n_q\in\mathcal{CN}(n_q;0,\rho(1-\rho))$ denotes
the quantization noise.
Notice that $\mathcal Q(x)=x$ means that no quantization effect is incorporated.


Then, from the above equation, we can know that the statistics
of the virtual channel $\mathbf{\widetilde h}_{k,m}$ can be achieved
through capturing the model parameter set $\boldsymbol\Xi_k=\{\alpha_k,\boldsymbol\Lambda_k\}$.
Moreover, once $\boldsymbol\Lambda_k$ is obtained,
we can obtain the non-zero supporting vector of $\mathbf{\tilde h}_{k,m}$,
 divide the users into different spatial groups,
and track $\mathbf {\tilde h}_{k,m}$.
Thus, in next section, we will resort to the damped Gaussian GAMP scheme with low complexity to
learn the prior model parameter $\boldsymbol\Xi_k$ and achieve the supporting vector of $\mathbf {\tilde h}_{k,m}$.

\section{Learning the Sparse Virtual Channel Model Parameters through Downlink Training}

Following most standards\cite{yang2013enhanced,larsson2001preamble}, we can fix one long training period called preamble along
the downlink to learn the model parameter set $\boldsymbol\Xi_k$. Without loss of generality, we use $M$ channel blocks. During
the $m$-th block, the BS transmits the $N\times P$ training matrix $\mathbf X_m$  with ${\mathbf X_m}^H \mathbf X_m = \frac{\sigma_p^2 \mathbf I_P}{P}$  to all the users, where $\sigma_p^2$ is
	the training power. Then, within the $m$-th block, the received training signal at user $k$ before ADC can be collected into a $P\times 1$ vector as
\begin{align}
\mathbf q_{k,m}=\mathbf X_m^T\mathbf h_{m,k}+\mathbf n_{k,m}=\mathbf X_m^T\mathbf F^H_N\mathbf{\widetilde h}_{k,m}+\mathbf n_{k,m},\label{eq:y_km}
\end{align}
where $\mathbf n_{k,m}$ is the independent additive white Gaussian noise vectorp with elements distributed as i.i.d. $\mathcal {CN}(0, \sigma_n^2)$, and $\sigma_n^2$ is assumed known.
Correspondingly, the quantization sample out of the ADC with respect to $\mathbf q_{k,m}$ at the receiver can be
written as
\begin{align}
\mathbf y_{k,m}=\mathcal Q(\mathbf q_{k,m}).
\end{align}

Let us define
  $N M\times 1$ vectors $\mathbf{\tilde h}_k=\left[\mathbf{\tilde h}_{k,1}^T,\mathbf{\tilde h}_{k,2}^T,\ldots,\mathbf{\tilde h}_{k,M}^T\right]^T$,
$\mathbf{r}_k=\left[\mathbf{r}_{k,1}^T,\mathbf r_{k,2}^T,\ldots,\mathbf r_{k,M}^T\right]^T$,
and   $PM\times 1$ vectors $\mathbf y_{k}=[\mathbf y_{k,1}^T,\mathbf y_{k,2}^T,\ldots,\mathbf y_{k,M}^T]^T$,
$\mathbf n_k=[\mathbf n_{k,1}^T,\mathbf n_{k,2}^T,\ldots,\mathbf n_{k,M}^T]^T$ for further use.
Obviously, through the downlink training, different users can independently learn their prior model parameters.
Thus, {in the following}, we will omit the user index $k$ for notational simplicity.

\subsection{Problem Formulation}
The learning objective  is to estimate the best fitting parameters set
$\boldsymbol\Xi$ with
the given observation vector $\mathbf y$.
Theoretically, the   ML estimator for $\boldsymbol\Xi$
can be formulated as
\begin{align}
\boldsymbol{\hat\Xi}=&
\arg \max_{1\geq \alpha \geq 0,~\lambda_{p}\geq 0}
\ln p(\mathbf y,\mathbf{\tilde h};\boldsymbol\Xi).
\end{align}
where $ p(\mathbf y,\mathbf{\tilde h};\boldsymbol\Xi)$ is the joint
PDF of $\mathbf{y}$ and $\mathbf{\tilde h}$
with given $\boldsymbol\Xi$. Obviously, such estimator involves  all possible combinations of the  $\mathbf{\tilde h}$ and is not feasible to directly achieve the ML solution due to its high dimensional search. Nonetheless,  one alternative method is to search
the solution iteratively   via the EM algorithm.

%
%
%
%

\subsection{the Low-complex Damped Gaussian GAMP based \text{EM} }

The EM algorithm
iteratively produces a sequence of ${\boldsymbol\Xi}^{(l)}, l = 1,2,\cdots $,
and each iteration is divided into two steps:

{\textbf{$\bullet$ Expectation step (E-step)}}
\begin{align}
Q\left({\boldsymbol\Xi},\widehat{\boldsymbol\Xi}^{(l-1)}\right)=&\mathbb E_{\mathbf{\tilde h}|\mathbf{ y}; \widehat{\boldsymbol\Xi}^{(l-1)}}
\Bigg\{ \ln p(\mathbf y,\mathbf{\tilde h};\boldsymbol\Xi)\Bigg\}.\label{eq:Q_Xi}
\end{align}

{\textbf{$\bullet$ Maximization step (M-step)}}
\begin{align}
{\widehat{\boldsymbol\Xi}}^{(l)}=&\arg\max_{{\boldsymbol\Xi}} Q\left({\boldsymbol\Xi},\widehat{\boldsymbol\Xi}^{(l-1)}\right).
\end{align}

\begin{figure}[!t]
	\centering
	\includegraphics[width=4.6in]{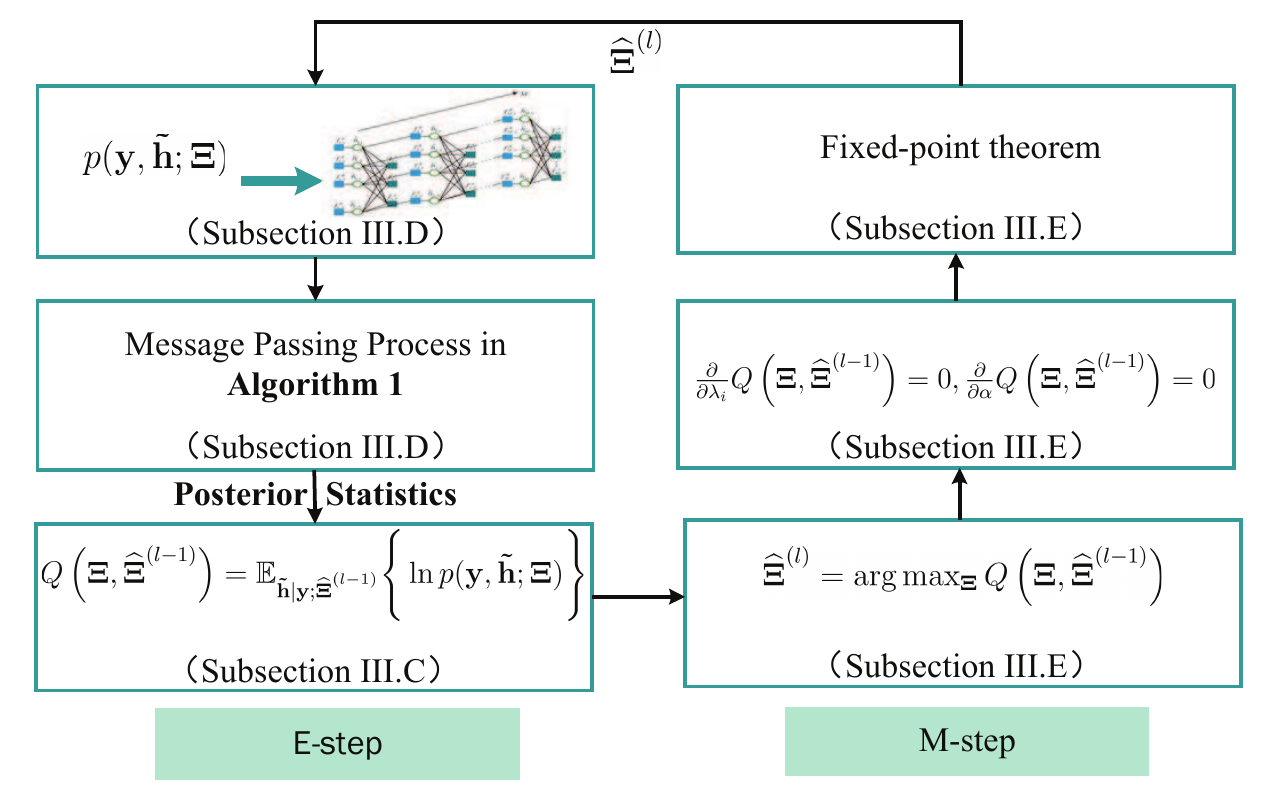}
	\caption{The  block diagram of the proposed  model parameter learning algorithm.}
	\label{fig:EM_DIAG}
\end{figure}
 {
During the iteration $l$, the E-step  is to derive the objective function $Q\big({\boldsymbol\Xi},\hat{\boldsymbol\Xi}^{(l-1)}\big)$ as  the expectation of  $p(\mathbf{ y}, \tilde{\mathbf{h}};\hat{\boldsymbol\Xi} )$ over $\tilde{\mathbf{h}}$ by setting $\boldsymbol\Xi$ as the  estimated model parameters $\hat{\boldsymbol\Xi}^{(l-1)}$ in the previous iteration (Subsection III.C). Specifically, the factor graph and the GAMP algorithms are used to compute the  desired posterior statistics (Subsection III.D).
The M-step  is to find the new estimation ${\boldsymbol\Xi^{(l)}}$ by
maximizing $Q\big({\boldsymbol\Xi},\hat{\boldsymbol\Xi}^{(l-1)}\big)$ (Subsection III.E).
In order to clearly describe the proposed model
parameters learning algorithm, we present its block diagram
in  \figurename { \ref{fig:EM_DIAG}}.
}

\subsection{ {Expectation step}}
In this subsection, we will   derive the  objective functions in
(\ref{eq:Q_Xi}).
Since the received samples $\mathbf y$ are known, the objective function $Q\left(\boldsymbol\Xi,\widehat{\boldsymbol\Xi}^{(l-1)}\right)$ can be
expressed as
\begin{align}
Q\left(\!\boldsymbol\Xi,\!\widehat{\boldsymbol\Xi}^{(l-1)}\!\right)
\!\!&=\!\mathbb E_{\mathbf{\tilde h}|\mathbf{ y}; \widehat{\boldsymbol\Xi}^{(l-1)}}
\Big\{ \!\ln p\Big(\!\mathbf{y}|\mathbf{\tilde h};\alpha\!\Big)\!\Big\}
\!+\!\mathbb E_{\mathbf{\tilde h}|\mathbf{ y}; \widehat{\boldsymbol\Xi}^{(l-1)}}
\Big\{ \ln p\Big(\mathbf{\tilde h}|\boldsymbol\lambda;\alpha\Big)\Big\}
\!\!+\!\mathbb E_{\mathbf{\tilde h}|\mathbf{ y}; \widehat{\boldsymbol\Xi}^{(l-1)}}
\Big\{\! \ln p\Big(\!\boldsymbol\lambda;\alpha\!\Big)\!\Big\},\notag\\
&=\sum_{m=1}^M
\mathbb E_{\mathbf{\tilde h}|\mathbf{ y}; \widehat{\boldsymbol\Xi}^{(l-1)}}
\Big\{ \ln p\Big(\mathbf{y}_m|\mathbf{\tilde h}_m;\alpha\Big)\Big\}
\!+\!\sum_{m=2}^M
\mathbb E_{\mathbf{\tilde h}|\mathbf{ y}; \widehat{\boldsymbol\Xi}^{(l-1)}}
\Big\{ \ln p\Big(\mathbf{\tilde h}_m|\mathbf{\tilde h}_{m-1},\boldsymbol\lambda;\alpha\Big)\Big\}\notag\\
&\kern 10pt+
\mathbb E_{\mathbf{\tilde h}|\mathbf{ y}; \widehat{\boldsymbol\Xi}^{(l-1)}}
\Big\{ \ln p\Big(\mathbf{\tilde h}_1,\boldsymbol\lambda;\alpha\Big)\Big\}
\!+\!\mathbb E_{\mathbf{\tilde h}|\mathbf{ y}; \widehat{\boldsymbol\Xi}^{(l-1)}}
\Big\{ \ln p\Big(\boldsymbol\lambda;\alpha\Big)\Big\}.
\label{eq:Q_function}
\end{align}

Under both the quantization and un-quantization case listed in the above section,
it can be verified  that the
conditional PDF $p\Big(\mathbf{y}_m|\mathbf{\tilde h}_m;\alpha\Big)$
is not related with the parameter set $\boldsymbol\Xi$.
Furthermore, with (\ref{eq:state}), it can be checked that
\begin{align}
&p\Big(\mathbf{\tilde h}_m|\mathbf {\tilde h}_{m-1}, \boldsymbol\lambda;\alpha\Big)
\sim \mathcal{CN}\Big(\mathbf{\tilde h}_{m};\alpha\mathbf{\tilde h}_{m-1},
(1-\alpha^2){\boldsymbol\Lambda}\Big),\\
&p\Big(\mathbf{\tilde h}_1|\boldsymbol\lambda;\alpha\Big)
\sim \mathcal{CN}\Big(\mathbf{\tilde h}_{1};\mathbf 0,{\boldsymbol\Lambda}\Big).
\label{eq:hm_con_alpha_PDF}
\end{align}
Plugging (\ref{eq:hm_con_alpha_PDF}) into (\ref{eq:Q_function}) and
taking some reorganizations, we can obtain

\begin{align}
Q\!\!\left(\!\boldsymbol\Xi,\!\widehat{\boldsymbol\Xi}^{(l-1)}\!\right)
\!\!&=\!\!\!\sum_{m=2}^M \!\!\Big[\!\frac{2\alpha}{1\!-\!\!\alpha^2}\tr\!\!\Big(\!\Re\Big\{\!{\boldsymbol\Lambda}^{-1}\mathbb E\Big\{\!
\mathbf{\tilde h}_{m-1}\mathbf{\tilde h}_{m}^H\!\Big|\!\mathbf y,\widehat{\boldsymbol\Xi}^{(\!l-\!1)}\!\Big\}\!\Big\}\!\Big)\!\!-\!\!\frac{\alpha^2}{1\!-\!\alpha^2}
\tr\!\!\Big(\!{\boldsymbol\Lambda}^{-1}\mathbb E\!\Big\{\!
\mathbf{\tilde h}_{m\!-\!1}\!\mathbf{\tilde h}_{m\!-\!1}^H\Big|\mathbf y,\widehat{\boldsymbol\Xi}^{(l-1)}\!\Big\}\!\Big)\!\Big]\notag\\
&-\sum_{m=2}^M\!\frac{1}{1\!-\!\alpha^2}\tr\Big({\boldsymbol\Lambda}^{-1}\mathbb E\Big\{
\mathbf{\tilde h}_{m}\mathbf{\tilde h}_{m}^H\Big|\mathbf y,\widehat{\boldsymbol\Xi}^{(l-1)}\Big\}\Big)
-\tr\Big({\boldsymbol\Lambda}^{-1}\mathbb E\Big\{
\mathbf{\tilde h}_{1}\mathbf{\tilde h}_{1}^H\Big|\mathbf y,\widehat{\boldsymbol\Xi}^{(l-1)}\Big\}\Big)\notag\\
&-(M-1)N\ln(1-\alpha^2)-M\ln|\boldsymbol\Lambda|
+\ln p(\boldsymbol\lambda;\alpha)+C,
\label{eq:Q_fuct_ex}
\end{align}
where $C$ is the items not related with $\boldsymbol\Xi$.

From (\ref{eq:Q_fuct_ex}),
it can be found that $Q\left(\boldsymbol\Xi,\hat{\boldsymbol\Xi}^{{(l-1)}}\right)$  is dependent on two posterior statistics,
i.e.,
$\mathbb E\left\{\mathbf{\tilde h}_{m}\mathbf{\tilde h}_{m}^H|
\mathbf y,\boldsymbol{\hat\Xi}^{(l-1)}\right\}$, and $\mathbb E\left\{\mathbf{ \tilde h}_{m-1}\mathbf{\tilde h}_{m}^H|
\mathbf y,\boldsymbol{\hat\Xi}^{(l-1)}\right\}$.
 Now, we turn to the calculations of these terms. Before calculating posterior statistics,
 let us define the following notations for further use
\begin{align}
\mathbf{\widehat{\tilde h} }_{m}^{(l)} & \!\stackrel{\vartriangle}{=}\! \mathbb E\left\{\!\mathbf{ \tilde h}_{m}|\mathbf y ,\boldsymbol{\hat\Xi}^{(l-1)}\!\right\},\kern 5pt
\boldsymbol\Theta_{m}^{(l)}\!\stackrel{\vartriangle}{=} \! \mathbb E\left\{\!\mathbf{\tilde h}_{m}\mathbf{\tilde h}_{m}^H|\mathbf y,\boldsymbol{\hat\Xi}^{(l-1)}\!\right\},\kern 5pt
\boldsymbol\Pi_{m-1,m}^{(l)}\!\stackrel{\vartriangle}{=}\!\mathbb E\left\{\!\mathbf{\tilde h}_{m-1}\mathbf{ \tilde h}_{m}^H|\mathbf y,\boldsymbol{\hat\Xi}^{(l-1)}\!\right\}.
\end{align}

\subsection{Deriving the Posterior Statistics with GAMP}

With given
$\mathbf y$ and $\boldsymbol{\hat\Xi}^{(l-1)}$, our objective is to infer
the posterior statistics $\widehat{\mathbf{\tilde{ h} }}_{m}^{(l)}$,
$\boldsymbol\Theta_{m}^{(l)}$, and $\boldsymbol\Pi_{m-1,m}^{(l)}$ under
the state-space model described by the following state equation in \eqref{eq:dss_state} and measurement equation in  \eqref{eq:dss_measure}:
\begin{align}
\mathbf{\tilde h}_{m}&= \hat{\alpha}^{(l-1)}  \mathbf{\tilde h}_{m-1}+  \sqrt{1-[\hat{\alpha}^{(l-1)}]^2} \boldsymbol\upsilon_{m},\label{eq:dss_state}\\
\mathbf y_m&=\mathcal{Q}\Big\{\underbrace{\mathbf X_m^T\mathbf F_N^H}_{\mathbf B_m}\mathbf{\tilde h}_{m}+\mathbf n_m\Big\},~m=1,2,\ldots,M,\label{eq:dss_measure}
\end{align}
where   matrix $\mathbf B_m$ is defined in \eqref{eq:dss_measure}, and $\boldsymbol\upsilon_{m}  \sim\mathcal{CN}(0,\boldsymbol \Lambda^{(l-1)})$.
With the Bayes rule,  the posterior joint probability density function
can be computed as
\begin{align}
p(\mathbf{\tilde h}|\mathbf y;\widehat{\boldsymbol\Xi}^{(l-1)})=\frac{
p\Big(\mathbf y|\mathbf{\tilde h};\widehat{\boldsymbol\Xi}^{(l-1)}\Big)
p\Big(\mathbf{\tilde h};\widehat{\boldsymbol\Xi}^{(l-1)}\Big)}{p\Big(\mathbf y;\widehat{\boldsymbol\Xi}^{(l-1)}\Big)}.
\label{eq:joint_post_PDF}
\end{align}
However, it is intractable to directly compute the desired posterior statistics, which is because of the
high-dimensional integrals over the marginal distributions. To avoid this obstacle, we will resort
to the factor graph and the message passing algorithms. First,
the posterior joint PDF in (\ref{eq:joint_post_PDF}) can be factorized as
\begin{align}
p(\mathbf{\tilde h}|\mathbf y;\widehat{\boldsymbol\Xi}^{(l-1)})\varpropto
&\prod_{m=1}^M\Big\{\prod_{p=1}^P  f^A_{m,p}(\mathbf {\tilde h}_{k,m})
\prod_{i=1}^N f^B_{m,i}(\tilde h_{m,i},\tilde h_{m-1,i})\Big\},\label{eq:ppost_PDF_fact}
\end{align}
where
$
f^A_{m,p}(\mathbf {\tilde h}_{k,m}) =p\Big(y_{m,p}|z_{m,p};\widehat{\boldsymbol\Xi}^{(l-1)}\Big),
f^B_{m,i}(\tilde h_{m,i},\tilde h_{m-1,i}) = p\Big(\tilde h_{m,i}|\tilde h_{m-1,i};\widehat{\boldsymbol\Xi}^{(l-1)}\Big)
$,$z_{m,p}=[\mathbf B_m]_{p,:}\mathbf{\tilde h}_m$,
  $m=1,2,\ldots,M$, $p=1,2,\ldots,P$, and $i=1,2,\ldots,N$;
 The explicit expressions of $p\Big(y_{m,p}|z_{m,p};\widehat{\boldsymbol\Xi}^{(l-1)}\Big)$ and $p\Big(\tilde h_{m,i}|\tilde h_{m-1,i};\widehat{\boldsymbol\Xi}^{(l-1)}\Big) $
are presented     in Appendix A.

%

\begin{figure}[!t]
	\centering
	\includegraphics[width=4.6in]{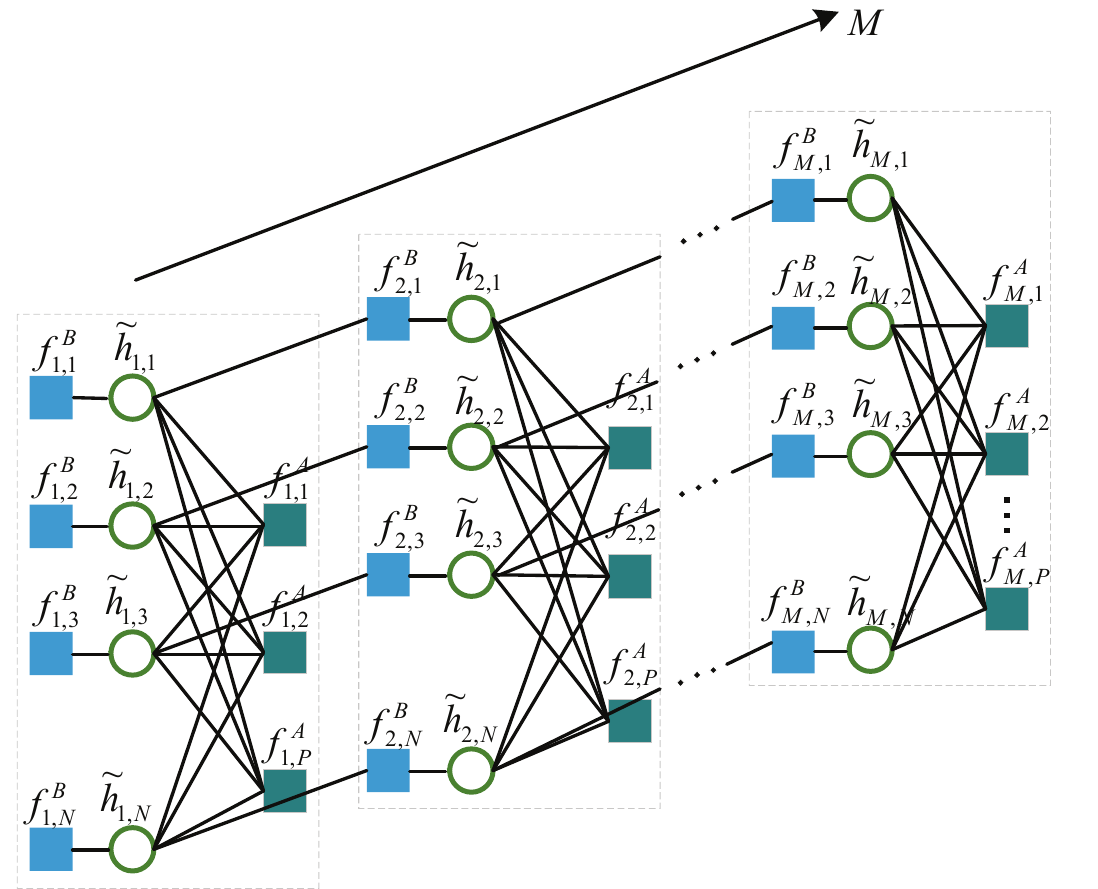}
	\caption{The Constructed Factor graph.}
	\label{fig:factor graph}
\end{figure}

 {Then, $p(\mathbf{\tilde h}|\mathbf y;\widehat{\boldsymbol\Xi}^{(l-1)})$
can be denoted with a factor
graph, as shown in Fig. \ref{fig:factor graph}}.
Obviously, there are  two kinds of function nodes, i.e.,
$f^A_{m,p}$, and
$f^B_{m,i}$, and the variable nodes, i.e., $\tilde h_{m,i}$, in Fig. \ref{fig:factor graph}.
 One specific variable node $x$ connects with
 the function nodes $f$, whose augments contain $x$.
 { Furthermore, 
for the function node $f$ and the variable node $x$, the messages from $f$ to $x$
 and from $x$ to $f$ are separately defined as $\vartheta_{f\rightarrow x}(x)$ and
 $\vartheta_{x\rightarrow f}(x)$, whose augment is $x$. With
 the belief propagation (BP) theory, we can obtain}
 \begin{align}
 \vartheta_{x\rightarrow f}(x)=\prod_{f^\prime\in \mathcal N(x)\slash f}\vartheta_{f^\prime \rightarrow x}(x),\kern 20pt
 \vartheta_{f\rightarrow x}(x)=\int_{\sim x}\Big(f(x)\prod_{x^\prime\in\mathcal{N}(f)\slash x}\vartheta_{x^\prime\rightarrow f}(x^\prime))\Big),\label{eq:sum_pro_rule}
 \end{align}
where the set $\mathcal N(x)$ collects all the neighbouring nods of the given node $x$
in one factor graph, and $\sim x$ possesses the same meaning with the same notation
in \cite{ kschischang2001factor}.

 {However, due to the presence of the cycles,
BP can not be directly applied for Fig. \ref{fig:factor graph}.
Nonetheless, the message scheduling and general approximate message propagation (GAMP) algorithms can be adopted
to effectively  approximate the  posterior distribution within the given allowable iterations.
Specially,
the message scheduling can be divided into
three steps, i.e., the forward message passing, the message exchanging, and the backward message passing.}
For clarity, we list the corresponding messages
in Table \ref{tab:message}.

\begin{table*}
     \begin{center}
       \caption{Different messages between nodes}
         \label{tab:message}

            \begin{tabularx}{\textwidth}{|X|l|l|} 
                \hline
                \textbf{Notations} & \textbf{Definitions}&\textbf{Values}\\
                \hline
               $\vartheta_{ f^A_{m,p}\to\tilde h_{m,i}}(\tilde h_{m,i})$ & belief from $f^A_{m,p}$ to $\tilde h_{m,i}$
               &$\mathcal{CN}\left(\tilde h_{m,i};\mu_{ f^A_{m,p}\to\tilde h_{m,i}}, \nu_{ f^A_{m,p}\to\tilde h_{m,i}}\right)$\\
               \hline
                 $\vartheta_{ f^B_{m,i}\to\tilde h_{m,i}}(\tilde h_{m,i})$ & belief from $f^B_{m,i}$ to $\tilde h_{m,i}$
               &$\mathcal{CN}\left(\tilde h_{m,i};\mu_{ f^B_{m,i}\to\tilde h_{m,i}}, \nu_{ f^B_{m,i}\to\tilde h_{m,i}}\right)$\\
               \hline
                 $\vartheta_{ f^B_{m+1,i}\to\tilde h_{m,i}}(\tilde h_{m,i})$ & belief from $f^B_{m+1,i}$ to $\tilde h_{m,i}$
               &$\mathcal{CN}\left(\tilde h_{m,i};\mu_{ f^B_{m+1,i}\to\tilde h_{m,i}}, \nu_{ f^B_{m+1,i}\to\tilde h_{m,i}}\right)$\\\hline
                            $\prod\limits_{p=1}^P\vartheta_{ f^A_{m,p}\to\tilde h_{m,i}}(\tilde h_{m,i})$ &
                             the sum product belief to $\tilde h_{m,i}$
               &$\mathcal{CN}\left(\tilde h_{m,i};\bar\mu_{ f^A_{m,:}\to\tilde h_{m,i}}, \bar\nu_{ f^A_{m,:}\to\tilde h_{m,i}}\right)$\\
               \hline
                \end{tabularx}
             \end{center}
           \end{table*}

\begin{figure}[!t]
	\centering
	\includegraphics[width=4.5in]{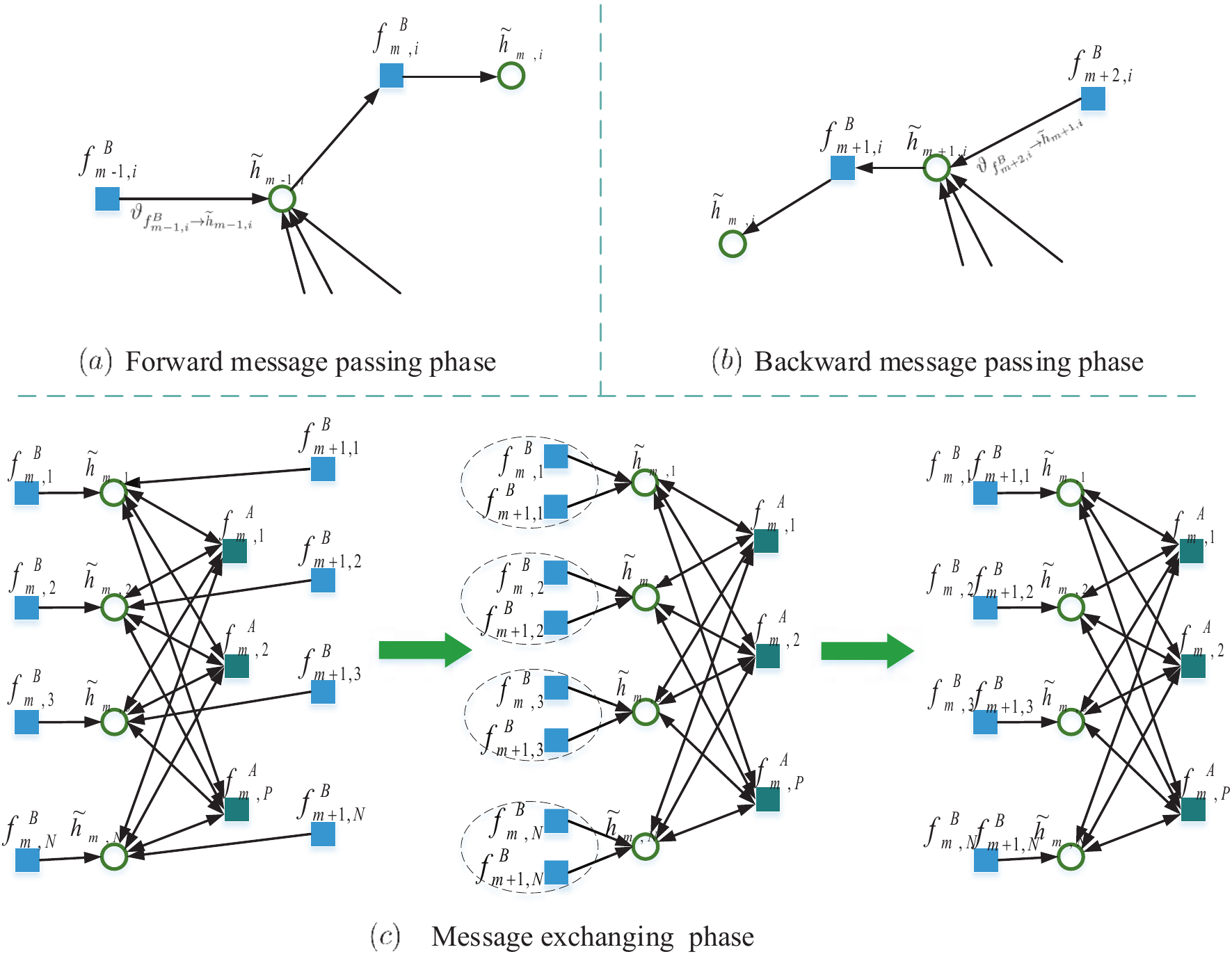}
	\caption{(a)The sub-factor-graph for the forward message passing into the mth time blockp. (b)The sub-factor-graph for the message exchanging within the $m$th time block. (c)The sub-factor-graph for the backward message passing into the mth time blockp.}
	\label{fig:Message_passing_phases}
\end{figure}
\subsubsection{\textbf{The belief updating for the forward message passing into the $\boldsymbol m$-th time block}}


Within this step, the beliefs are passed from the variable nodes
$\tilde h_{m-1,i}$  into the $m$th time block through the function nodes $f^B_{m,i}$, and the believes
$\vartheta_{f_{m,i}^B\to \tilde h_{m,i}}$ will be updated in the sub-graph Fig. \ref{fig:Message_passing_phases}(a)
\cite{ziniel2012generalized}.
Thus, for $2\le m\le M$,
 {with (\ref{eq:ppost_PDF_fact}), (\ref{eq:sum_pro_rule}) and Table. \ref{tab:message}}, we can obtain

\begin{align}
&\vartheta_{f_{m,i}^B\to \tilde h_{m,i}}\! \propto\!\int_{\tilde h_{m-1,i}} \!\!f_{m,i}^B\vartheta_{\tilde h_{m-1,i}\to f_{m,i}^B} d \tilde h_{m-1,i}\!=\!\int_{\tilde h_{m-1,i}} f_{m,i}^B\vartheta_{f_{m-1,i}^B\to \tilde h_{m-1,i}}
\prod_{P=1}^P\vartheta_{f_{m-1,p}^A\to \tilde h_{m-1,i}}   d \tilde h_{m-1,i}\notag\\
&\propto
\int_{\tilde h_{m-1,i}}\mathcal{CN}\left(\tilde h_{m-1,i};\mu_{f_{m-1,i}^B\to \tilde h_{m-1,i}}, \nu_{f_{m-1,i}^B\to \tilde h_{m-1,i}}\right)
\mathcal{CN}\left(\tilde h_{m-1,i};{\bar\mu}_{f_{m-1,:}^A\to \tilde h_{m-1,i}}, \bar{\nu}_{f_{m-1,:}^A\to \tilde h_{m-1,i}}\right)\notag\\
&\times\mathcal{CN}\left(\tilde h_{m,i};
\hat\alpha^{(l-1)}\tilde h_{m-1,i},\Big(1-\Big[\hat\alpha^{(l-1)}\Big]^2\Big)\hat\lambda_{i}^{(l-1)}\right)
d \tilde h_{m-1,i}\notag\\
&=\mathcal{CN}\left(\tilde h_{m,i};\mu_{f_{m,i}^B\to \tilde h_{m,i}}, \nu_{f_{m,i}^B\to \tilde h_{m,i}}\right),
\label{eq:vartheta_fB_h }
\end{align}
 where
 \begin{align}
 \mu_{f_{m,i}^B\to \tilde h_{m,i}}=&
 \hat\alpha^{(l-1)}\left(\frac{\bar\mu_{f_{m-1,:}^A\to \tilde h_{m-1,i}}}{\bar\nu_{f_{m-1,:}^A\to \tilde h_{m-1,i}}}+
 \frac{\mu_{f_{m-1,i}^B\to \tilde h_{m-1,i}}}{\nu_{f_{m-1,i}^B\to \tilde h_{m-1,i}}}\right)
 \frac{\bar\nu_{f_{m-1,:}^A\to \tilde h_{m-1,i}}\nu_{f_{m-1,i}^B\to \tilde h_{m-1,i}}}{\bar\nu_{f_{m-1,i}^A\to \tilde h_{m-1,i}}+\nu_{f_{m-1,i}^B\to \tilde h_{m-1,i}}},
  \label{eq:mup_fB_h }\\
  \nu_{f_{m,i}^B\to \tilde h_{m,i}}=& \left[\hat\alpha^{(l-1)}\right]^2\frac{\bar\nu_{f_{m-1,:}^A\to \tilde h_{m-1,i}}\nu_{f_{m-1,i}^B\to \tilde h_{m-1,i}}}{\bar\nu_{f_{m-1,:}^A\to \tilde h_{m-1,i}}+\nu_{f_{m-1,i}^B\to \tilde h_{m-1,i}}}+\left(1-\left[\hat\alpha^{(l-1)}\right]^2\right)\hat\lambda_{i}^{(l-1)},
  \label{eq:nu_fB_h }
\end{align}
 and the property $\prod\limits_{n}\mathcal{CN}(r;\mu_n,\nu_n)\propto\mathcal{CN}\left(r;\frac{\sum\limits_n\mu_n/\nu_n}{\sum\limits_n 1/\nu_n}, \frac{1}{\sum\limits_n1/\nu_n}\right)$ is utilized in the above derivation.
Specially, at $M=1$, we have $\vartheta_{f_{1,i}^B\to \tilde h_{1,i}}=\mathcal{CN}(\tilde h_{1,i};
\mu_{f_{1,i}^B\to \tilde h_{1,i}}, \nu_{f_{1,i}^B\to \tilde h_{1,i}})$,
$\mu_{f_{1,i}^B\to \tilde h_{1,i}}=0$, and $\nu_{f_{1,i}^B\to \tilde h_{1,i}}=\hat\lambda_{1}^{(l-1)}$.

\subsubsection{\textbf{The belief updating for the message exchanging within
the $\boldsymbol m$-th time block}}

In this step, we will obtain the estimation of $\tilde h_{m,i}$ through exchanging
information within the $m$th time block, where \textcolor{black}{$\mathbf y_m$, and
$\vartheta_{ f^B_{m+1,i}\to\tilde h_{m,i}}$, $\vartheta_{ f^B_{m,i}\to\tilde h_{m,i}}$
are utilized.}
\textcolor{black}{With the left sub-figure of Fig. \ref{fig:Message_passing_phases}(c) and Table.~\ref{tab:message}},
the sum product belief from both $f^B_{m+1,i}$
and $f^B_{m,i}$ to $\tilde h_{m,i}$ can be written as
\begin{align}
\vartheta_{\{f^B_{m+1,i}, f^B_{m,i}\}\to\tilde h_{m,i}}
\propto&\mathcal{CN}\left(\tilde h_{m,i};\mu_{ f^B_{m+1,i}\to\tilde h_{m,i}}, \nu_{ f^B_{m+1,i}\to\tilde h_{m,i}}\right)
\mathcal{CN}\left(\tilde h_{m,i};\mu_{ f^B_{m,i}\to\tilde h_{m,i}}, \nu_{ f^B_{m,i}\to\tilde h_{m,i}}\right)\notag\\
\propto&\mathcal{CN}\left(\tilde h_{m,i};\mu_{\{f^B_{m+1,i}, f^B_{m,i}\}\to\tilde h_{m,i}},\nu_{\{f^B_{m+1,i}, f^B_{m,i}\}\to\tilde h_{m,i}}\right),\label{eq:vartheta_neigh_fB_h}
\end{align}
\begin{align}
\text{where}~~~~~~~~~~~\nu_{\{f^B_{m+1,i}, f^B_{m,i}\}\to\tilde h_{m,i}}=&\frac{\nu_{f_{m,i}^B\to \tilde h_{m,i}}\nu_{f_{m+1,i}^B\to \tilde h_{m,i}}}{\nu_{f_{m,i}^B\to \tilde h_{m,i}}+\nu_{f_{m+1,i}^B\to \tilde h_{m,i}}}, \label{eq:nu_neigh_fB_h}\\
\mu_{\{f^B_{m+1,i}, f^B_{m,i}\}\to\tilde h_{m,i}}=&\nu_{\{f^B_{m+1,i}, f^B_{m,i}\}\to\tilde h_{m,i}}
 \left(\frac{\mu_{f_{m,i}^B\to \tilde h_{m,i}}}{\nu_{f_{m,i}^B\to \tilde h_{m,i}}}+
 \frac{\mu_{f_{m+1,i}^B\to \tilde h_{m,i}}}{\nu_{f_{m+1,i}^B\to \tilde h_{m,i}}}\right).
 \label{eq:mu_neigh_fB_h}~~~~~
 \end{align}
 With this observation, we can treat the nodes $\{f^B_{m+1,i},f^B_{m,i}\}$
 together, and obtain    Fig. \ref{fig:Message_passing_phases}(c),
 which is the same to the factor graph for the GAMP \cite{al2018gamp}.
 In fact, if we only consider the right sub-figure of   Fig. \ref{fig:Message_passing_phases}(c),
 the MMSE estimation of $\tilde h_{m,i}$ can be solved through the GAMP with respect
 to the observation model
 \begin{align}
 \mathbf y_m=\mathcal Q\{\mathbf z_m+\mathbf n_m\},
 \label{eq:GAMP_mea}
 \end{align}
 where
 $\mathbf z_m=[z_{m,1},z_{m,2},\ldots,z_{m,P}]^T$,
 and the
 prior knowledge about $\tilde h_{m,i}$ can be expressed as
 \begin{align}
 p(\tilde h_{m,i})=\mathcal{CN}\left(\tilde h_{m,i};\mu_{\{f^B_{m+1,i}, f^B_{m,i}\}\to\tilde h_{m,i}},\nu_{\{f^B_{m+1,i}, f^B_{m,i}\}\to\tilde h_{m,i}}\right). \label{eq:GAMP_prior}
 \end{align}

\subsubsection{\textbf{The backward message passing into $\boldsymbol m$-th time block.}}
Within this step, the believes about variable node  $\tilde h_{m+1,i}$
will be passed into the $m$th time block in the backward manner within
Fig. \ref{fig:Message_passing_phases}(b), where
the related belief is $\vartheta_{f_{m+1,i}^B\to \tilde h_{m,i}}$, respectively.
Following the similar methods, we can calculate
$\vartheta_{f_{m+1,i}^B\to \tilde h_{m,i}}$ as
\begin{align}
&\vartheta_{f_{m+1,i}^B\to \tilde h_{m,i}}\notag\\
&\propto\int_{\tilde h_{m+1,i}} f_{m+1,i}^B\vartheta_{\tilde h_{m+1,i}\to f_{m+1,i}^B} d \tilde h_{m+1,i}=\int_{\tilde h_{m+1,i}} f_{m+1,i}^B\vartheta_{f_{m+2,i}^B\to \tilde h_{m+1,i}}
\prod_{P=1}^P\vartheta_{f_{m+1,p}^A\to \tilde h_{m+1,i}}   d \tilde h_{m+1,i}\notag\\
&\propto
\int_{\tilde h_{m+1,i}}\mathcal{CN}\left(\tilde h_{m+1,i};\mu_{f_{m+2,i}^B\to \tilde h_{m+1,i}}, \nu_{f_{m+2,i}^B\to \tilde h_{m+1,i}}\right)
\mathcal{CN}\left(\tilde h_{m+1,i};{\bar\mu}_{f_{m+1,:}^A\to \tilde h_{m+1,i}}, \bar{\nu}_{f_{m+1,:}^A\to \tilde h_{m+1,i}}\right)\notag\\
&~~\times\mathcal{CN}\left(\tilde h_{m+1,i};
\hat\alpha^{(l-1)}\tilde h_{m,i},\Big(1-\Big[\hat\alpha^{(l-1)}\Big]^2\Big)\hat\lambda_{i}^{(l-1)}\right)
d \tilde h_{m+1,i}\notag\\
&=\mathcal{CN}\left(\tilde h_{m,i};\mu_{f_{m+1,i}^B\to \tilde h_{m,i}}, \nu_{f_{m+1,i}^B\to \tilde h_{m,i}}\right),
\label{eq:vartheta_fB_h_back}
\end{align}
 where
 \begin{align}
 \mu_{f_{m+1,i}^B\to \tilde h_{m,i}}=&
 \frac{1}{\hat\alpha^{(l-1)}}
 \left(\frac{\bar\mu_{f_{m+1,:}^A\to \tilde h_{m+1,i}}}{\bar\nu_{f_{m+1,:}^A\to \tilde h_{m+1,i}}}+
 \frac{\mu_{f_{m+2,i}^B\to \tilde h_{m+1,i}}}{\nu_{f_{m+2,i}^B\to \tilde h_{m+1,i}}}\right)
 \frac{\bar\nu_{f_{m+1,:}^A\to \tilde h_{m+1,i}}\nu_{f_{m+2,i}^B\to \tilde h_{m+1,i}}}{\bar\nu_{f_{m+1,i}^A\to \tilde h_{m+1,i}}+\nu_{f_{m+2,i}^B\to \tilde h_{m+1,i}}},
 \label{eq:mu_fB_h_back}\\
  \nu_{f_{m+1,i}^B\to \tilde h_{m,i}}=& \frac{1}{\left[\hat\alpha^{(l-1)}\right]^2}\left(\frac{\bar\nu_{f_{m+1,:}^A\to \tilde h_{m+1,i}}\nu_{f_{m+2,i}^B\to \tilde h_{m+1,i}}}{\bar\nu_{f_{m+1,:}^A\to \tilde h_{m+1,i}}+\nu_{f_{m+2,i}^B\to \tilde h_{m+1,i}}}+\left(1-\left[\hat\alpha^{(l-1)}\right]^2\right)\hat\lambda_{i}^{(l-1)}\right),
   \label{eq:nu_fB_h_back}
\end{align}
$m=1,2,\ldots, M-1$. Furthermore, we can set
$\vartheta_{f_{M+1,i}^B\to \tilde h_{M,i}}=\mathcal{CN}(\tilde h_{M,i};0,+\infty)$.

\begin{algorithm}[!thp]
\setstretch{1.22}
    \caption{The Message Passing Process for the expectation step of the $l$th EM iteration} 
    \begin{algorithmic}[1]\label{alg:GAMP_EM}
    \REQUIRE: Matrix $\mathbf{B}_m$, scalar estimation functions $g_{s}$ and $g_{\tilde h}$,and damping constants $\theta_{s}$,$\theta_{\tilde h}\in(0,1]$.
        \STATE k=0, Initialize $\boldsymbol{\nu}_{{\tilde h},m}^{(k=0)}>\text{diag}\Big(    \boldsymbol\Theta_m^{(l-1)}-\widehat{\mathbf{\tilde h}}_m^{(l-1)}\Big(\widehat{\mathbf{\tilde h}}_m^{(l-1)}\Big)^H\Big),\mathbf{\tilde h}_m^{(k=0)}=\widehat{\mathbf{\tilde h}}_m^{(l-1)}$,
        $\mathbf s_m^{(k=0)}=\widehat{\mathbf s}_m^{(l-1)}$.
        \REPEAT
         \STATE Implement the forward message passing into the $m$th time block, $i=1$.
        \REPEAT
        \STATE $k=k+1$
        \STATE $i=i+1$, $m=1$, $\mu_{f_{1,i}^B\to \tilde h_{1,i}}=0$, $\nu_{f_{1,i}^B\to \tilde h_{1,i}}=\hat\lambda_{1}^{(l-1)}$.
         \REPEAT
            \STATE  $m=m+1$.
            \STATE  $\mu_{f_{m,i}^B\to \tilde h_{m,i}}=
                     \hat\alpha^{(l-1)}\left(\frac{\bar\mu_{f_{m-1,:}^A\to \tilde h_{m-1,i}}}{\bar\nu_{f_{m-1,:}^A\to \tilde h_{m-1,i}}}+\frac{\mu_{f_{m-1,i}^B\to \tilde h_{m-1,i}}}{\nu_{f_{m-1,i}^B\to \tilde h_{m-1,i}}}\right)\frac{\bar\nu_{f_{m-1,:}^A\to \tilde h_{m-1,i}}\nu_{f_{m-1,i}^B\to \tilde h_{m-1,i}}}{\bar\nu_{f_{m-1,i}^A\to \tilde h_{m-1,i}}+\nu_{f_{m-1,i}^B\to \tilde h_{m-1,i}}}$.
            \STATE $\nu_{f_{m,i}^B\to \tilde h_{m,i}}=\left[\hat\alpha^{(l-1)}\right]^2\frac{\bar\nu_{f_{m-1,:}^A\to \tilde h_{m-1,i}}\nu_{f_{m-1,i}^B\to \tilde h_{m-1,i}}}{\bar\nu_{f_{m-1,:}^A\to \tilde h_{m-1,i}}+\nu_{f_{m-1,i}^B\to \tilde h_{m-1,i}}}+\left(1-\left[\hat\alpha^{(l-1)}\right]^2\right)\hat\lambda_{i}^{(l-1)}$¡£
            \UNTIL $m=M$
        \UNTIL $i=N$
        \STATE Implement the message exchanging within
the $m$th time block, $m=1$
        \REPEAT
            \STATE $\mathbf{S}=\mathbf{B}_m.\mathbf B_m$(component-wise magnitude squared)
         \STATE 1./$\boldsymbol{\nu}_{p,m}^{(k)}=\mathbf{S}\boldsymbol{\nu}_{{\tilde h},m}^{(k)}$,\kern 20pt $\mathbf{p}_m^{(k)}=\mathbf{s}_m^{(k-1)}+\boldsymbol{\nu}_{p,m}^{(k)}.\mathbf{B}_m\mathbf{\tilde h}_m^{(k)}$.
         \STATE $\boldsymbol{\nu}_{s,m}^{(k)}=\boldsymbol{\nu}_{p,m}^{(k)}.g_{s}^{\prime}\left(\mathbf{p}_m^{{(k)}},\boldsymbol\nu_{p,m}^{(k)}\right)$,
         \kern 10pt $\mathbf s_m^{(k)}=(1-\theta_{s})\mathbf{s}_m^{(k-1)}+\theta_{s}g_{s}\left(\mathbf{p}_m^{(k)},\boldsymbol{\nu}_{p,m}^{(k)}\right)$.
         \STATE 1./${\boldsymbol\nu}_{r,m}^{(k)}=\mathbf{S}_m^{T}\boldsymbol{\nu}_{s,m}^{(k)}$,\kern 20pt $\mathbf{r}_m^{(k)}=\mathbf{\tilde h}_m^{(k)}-\boldsymbol{\nu}_{r,m}^{(k)}.\mathbf{B}_m^{H}\mathbf{s}_m^{(k)}$.
         \STATE $\boldsymbol{\tau}_{{\tilde h},m}^{(k+1)}=\boldsymbol{\nu}_{r,m}^{(k)}.g_{{\tilde h},m}^{\prime}(\mathbf{r}_m^{(k)},\boldsymbol{\nu}_{r,m}^{(k)})$,\kern 10pt $\mathbf{\tilde h}_m^{(k+1)}=(1-\theta_{\tilde h})\mathbf{\tilde h}_m^{{k}}+\theta_{\tilde h}g_{{\tilde h},m}(\mathbf{r}_m^{(k)},\boldsymbol{\nu}_{r,m}^{(k)})$.
         \STATE $m=m+1$.
        \UNTIL $m=M$
        \STATE for $i=1$ to $i=N$, $\bar\mu_{ f^A_{m,:}\to\tilde h_{m,i}}=[\mathbf r_{m}^{(k)}]_{i}$, $\bar\nu_{ f^A_{m,:}\to\tilde h_{m,i}}=[\boldsymbol\nu_{r,m}^{(k)}]_{i}$.
            \STATE Implement the backward message passing into the $m$th time block, $i=1$.
            \REPEAT
        \STATE $i=i+1$, $m=M$, $\mu_{f_{M+1,i}^B\to \tilde h_{M,i}}=0$, $\nu_{f_{M+1,i}^B\to \tilde h_{M,i}}=\mathcal{CN}(\tilde h_{M,i};0,+\infty)=\infty$.
         \REPEAT
            \STATE  $m=m-1$.
    \STATE  $\mu_{f_{m+1,i}^B\to \tilde h_{m,i}}=\frac{1}{\hat\alpha^{(l-1)}}
 \left(\frac{\bar\mu_{f_{m+1,:}^A\to \tilde h_{m+1,i}}}{\bar\nu_{f_{m+1,:}^A\to \tilde h_{m+1,i}}}+
 \frac{\mu_{f_{m+2,i}^B\to \tilde h_{m+1,i}}}{\nu_{f_{m+2,i}^B\to \tilde h_{m+1,i}}}\right)
 \frac{\bar\nu_{f_{m+1,:}^A\to \tilde h_{m+1,i}}\nu_{f_{m+2,i}^B\to \tilde h_{m+1,i}}}{\bar\nu_{f_{m+1,i}^A\to \tilde h_{m+1,i}}+\nu_{f_{m+2,i}^B\to \tilde h_{m+1,i}}}$,
 \STATE $\nu_{f_{m+1,i}^B\to \tilde h_{m,i}}= \frac{1}{\left[\hat\alpha^{(l-1)}\right]^2}\left(\frac{\bar\nu_{f_{m+1,:}^A\to \tilde h_{m+1,i}}\nu_{f_{m+2,i}^B\to \tilde h_{m+1,i}}}{\bar\nu_{f_{m+1,:}^A\to \tilde h_{m+1,i}}+\nu_{f_{m+2,i}^B\to \tilde h_{m+1,i}}}+\left(1-\left[\hat\alpha^{(l-1)}\right]^2\right)\hat\lambda_{i}^{(l-1)}\right)$,
            \UNTIL $m=1$
        \UNTIL $i=N$
        \UNTIL $k=K_{max}$
        \STATE Output the estimation results of the $l$th EM iteration, i.e.,
         $\widehat{\mathbf s}_m^{(l)}={\mathbf s}_m^{(K_{max})}$, $\widehat{\mathbf{\tilde h}}_m^{(l)}={\mathbf{\tilde h}}_m^{(K_{max}+1)}$, $
         \boldsymbol\Theta_m^{(l)}
         =\text{Diag}\Big(\boldsymbol{\tau}_{{\tilde h},m}^{(K_{max}+1)}\Big)+
         \widehat{\mathbf{\tilde h}}_m^{(l)}\Big(\widehat{\mathbf{\tilde h}}_m^{(l)}\Big)^H$, $i=1,2,\ldots, N$, $M=1,2,\ldots,M$.
    \end{algorithmic}
\end{algorithm}

Taking the above three message updating phases into consideration, we can listed
the detailed steps for the expectation step of the $l$th EM iteration in \textbf{Algorithm} \ref{alg:GAMP_EM}.
In this algorithm, the notations $\mathbf a.\mathbf b$ and $\mathbf a./\mathbf b$ denote the
component-wise multiplication and division, respectively. Furthermore,
the input scalar estimation function $g_{s}\left(\mathbf{p},\boldsymbol{\nu}_{\mathbf p}\right)$
and that for the output $g_{\tilde h}(\mathbf{r},\boldsymbol{\nu}_{\mathbf r})$
can be separately defined as

\begin{align}
\left[g_{s}\left(\mathbf{p},\boldsymbol{\nu}_{p}\right)\right]_n=&p_n-
\nu_{p,n}\frac{\int z_{m,n}p({y_{m,n}|z_{m,n}})\mathcal{CN}({z}_{m,n};\frac{p_n}{\nu_{p,n}},\frac{1}{\nu_{p,n}})
d{z}_{m,n}}{\int p({y_{m,n}|z_{m,n}})\mathcal{CN}({z}_{m,n};\frac{p_n}{\nu_{p,n}},\frac{1}{\nu_{p,n}})
d{z}_{m,n}},\label{eq:g_s_def}\\
\left[g_{{\tilde h}}\left(\mathbf{r},\boldsymbol{\nu}_{\mathbf r}\right)\right]_i=&
\frac{\int {\tilde h}_{m,i}p({\tilde h}_{m,i})\mathcal{CN}({\tilde h}_{m,i};r_i,\nu_{{r},i})
d{\tilde h}_{m,i}}{\int p({\tilde h}_{m,i})\mathcal{CN}({\tilde h}_{m,i};r_i,\nu_{{r},i})
d{\tilde h}_{m,i}},\label{eq:g_h_def}
\end{align}
where $\mathbf r=[r_1,r_2,\ldots,r_N]^T$,
$\boldsymbol\nu_{r}=[\nu_{r,1},\nu_{r,2},\ldots,\nu_{r,N}]^T$,
$\mathbf p=[p_1,p_2,\ldots,p_P]^T$, and
$\boldsymbol\nu_{p}=[\nu_{p,1},\nu_{p,2},\ldots,\nu_{p,P}]^T$.
In Appendix B, we  derive
$g_{s}\left(\mathbf{p},\boldsymbol{\nu}_{\mathbf p}\right)$, $g_{\tilde h}(\mathbf{r},\boldsymbol{\nu}_{\mathbf r})$
and their corresponding partial derivatives. For clarity, we show them in Table. \ref{tab:g_s_h},
 {where the explicit expressions of  $\Delta^{\Re}_{m,n}$,
 $\Delta^{\Im}_{m,n}$,
$\nabla^{\Re}_{m,n}$, $\nabla^{\Im}_{m,n}$,
$\Xi^{\Re}_{m,n}$, and $\Xi^{\Im}_{m,n}$
are presented in Appendix B.}

\begin{table*}
     \begin{center}
       \caption{The values of $g_{s}\left(\mathbf{p},\boldsymbol{\nu}_{\mathbf p}\right)$, $g_{\tilde h}(\mathbf{r},\boldsymbol{\nu}_{\mathbf r})$
and their partial derivatives under different quantization cases}
         \label{tab:g_s_h}

            \begin{tabularx}{\textwidth}{|X |l|} 
                \hline
                \textbf{The quantization Cases}&\textbf{The values of $g_{s}\left(\mathbf{p}\!,\!\boldsymbol{\nu}_{\mathbf p}\right)$, $g_{\tilde h}(\mathbf{r},\!\boldsymbol{\nu}_{\mathbf r})$
and their partial derivatives}\\
                \hline
                No quantization
                &
$\left[g_{\mathbf s}\left(\mathbf{p},\boldsymbol{\nu}_{\mathbf p}\right)\right]_n
=\frac{p_n-\nu_{p,n}{y_{m,n}}}{1+\nu_{p,n}{\sigma_n^2}}$,
$\left[g_{\mathbf s}^\prime\left(\mathbf{p},\boldsymbol{\nu}_{\mathbf p}\right)\right]_n=\frac{1}{1+\nu_{p,n}{\sigma_n^2}}$.
                \\
               \hline
                 Normal Quantization & $\left[g_{\mathbf s}\left(\mathbf{p},\boldsymbol{\nu}_{\mathbf p}\right)\right]_n\!=\!\!\frac{1}{2}\frac{\Delta^{\Re}_{m,n}}{\nabla^{\Re}_{m,n}}
+\frac{1}{2}\jmath\frac{\Delta^{\Im}_{m,n}}{\nabla^{\Im}_{m,n}}$,\\
&
$[g_{s}^\prime\left(\mathbf{p},\boldsymbol{\nu}_{\mathbf p}\right)]_n\!=\!
\frac{1}{4\nu_{p,n}\sqrt{\frac{1}{2}(\sigma_n^2\!+\!\frac{1}{\nu_{p,n}})}}
\left\{\frac{\Xi^{\Re}_{m,n}}{\nabla^{\Re}_{m,n}}\!+\!\frac{\Xi^{\Im}_{m,n}}{\nabla^{\Im}_{m,n}} \right\}
\!+\!\frac{1}{\nu_{p,n}}|\left[\!g_{s}\left(\!\mathbf{p},\boldsymbol{\nu}_{p}\!\right)\!\right]_n|^2$.
              \\\hline
                            PDQ case &
              $\left[g_{\mathbf s}\left(\mathbf{p},\boldsymbol{\nu}_{\mathbf p}\right)\right]_n
=\frac{(1-\rho)p_n-\nu_{p,n}y_{m,n}}{(1-\rho)+\nu_{p,n}[(1-\rho)\sigma_n^2+\rho]}$,
$\left[g_{\mathbf s}^\prime\left(\mathbf{p},\boldsymbol{\nu}_{\mathbf p}\right)\right]_n=\frac{1-\rho}{(1-\rho)+\nu_{p,n}[(1-\rho)\sigma_n^2+\rho]}$.
            \\
               \hline
               All the cases &
       $\left[g_{{\tilde h}}\left(\mathbf{r},\boldsymbol{\nu}_{\mathbf r}\right)\right]_i=
\frac{\mu_{\{f^B_{m+1,i}, f^B_{m,i}\}\to\tilde h_{m,i}}\nu_{{r},i}+r_i\nu_{\{f^B_{m+1,i}, f^B_{m,i}\}\to\tilde h_{m,i}}}{\nu_{\{f^B_{m+1,i}, f^B_{m,i}\}\to\tilde h_{m,i}}+\nu_{{r},i}}$,\\
&
$\left[g_{{\tilde h}}^\prime\left(\mathbf{r},\boldsymbol{\nu}_{\mathbf r}\right)\right]_i
=\frac{\nu_{\{f^B_{m+1,i}, f^B_{m,i}\}\to\tilde h_{m,i}}}{\nu_{\{f^B_{m+1,i}, f^B_{m,i}\}\to\tilde h_{m,i}}+\nu_{{r},i}}$.
  \\\hline
                \end{tabularx}
             \end{center}
           \end{table*}


\subsection{Maximization Step}

In this step, we will derive $\boldsymbol{\hat\Xi}^{(l)}$ through maximizing  $Q\big({\boldsymbol\Xi},\hat{\boldsymbol\Xi}^{(l-1)}\big)$  as
\begin{align}
\boldsymbol{\hat\Xi}^{(l)}=\arg\max_{{\boldsymbol\Xi}}\left\{Q\big({\boldsymbol\Xi},\hat{\boldsymbol\Xi}^{(l-1)}\big)\right\}.
\tag{P1}
\end{align}
Taking the derivative of (\ref{eq:Q_fuct_ex}) with respect to $\lambda_i$
and $\alpha$, we can obtain
\begin{align}
\frac{\partial}{\partial \lambda_i}Q\left(\boldsymbol\Xi,\widehat{\boldsymbol\Xi}^{(l-1)}\right)
=  &\sum_{m=2}^M \Big[-\frac{2\alpha}{1-\alpha^2}\frac{1}{\lambda_i^2}\Re\Big\{\Big[\boldsymbol\Pi_{m-1,m}^{(l)}
\Big]_{i,i}\Big\}+\frac{\alpha^2}{1-\alpha^2}
\frac{1}{\lambda_i^2}\Big[\boldsymbol\Theta_{m-1}^{(l)}\Big]_{i,i}\Big]\notag\\
&+\sum_{m=2}^M\frac{1}{1-\alpha^2}\frac{1}{\lambda_i^2}\Big[\boldsymbol\Theta_{m}^{(l)}\Big]_{i,i}
+\frac{1}{\lambda_i^2}\Big[\boldsymbol\Theta_{1}^{(l)}\Big]_{i,i}-M\frac{1}{\lambda_i}
+\frac{\partial\ln p(\boldsymbol\lambda)}{\partial\lambda_i},\label{eq:Q_der_lambda}
\end{align}

\begin{align}
\frac{\partial}{\partial \alpha}Q\left(\boldsymbol\Xi,\widehat{\boldsymbol\Xi}^{(l-1)}\right)
=&\sum_{m=2}^M \Big[\frac{2+2\alpha^2}{(1-\alpha^2)^2}\tr\Big(\Re\Big\{{\boldsymbol\Lambda}^{-1}\boldsymbol\Pi_{m-1,m}^{(l)}\Big\}\Big)
-\frac{2\alpha}{(1-\alpha^2)^2}
\tr\Big({\boldsymbol\Lambda}^{-1}\boldsymbol\Theta_{m-1}^{(l)}\Big)\Big]\notag\\
&-\sum_{m=2}^M\frac{2\alpha}{(1-\alpha^2)^2}\tr\Big({\boldsymbol\Lambda}^{-1}\boldsymbol\Theta_{m}^{(l)}\Big)
+(M-1)N\frac{2\alpha}{1-\alpha^2}. \label{eq:Q_der_alpha}
\end{align}

Theoretically, under the sparse Bayesian learning framework,
the non-informative prior is used for $p(\boldsymbol\lambda)$. Hence,
we can ignore the effect of $p(\boldsymbol\lambda)$ in the maximization step.
Correspondingly,   with fixed $\alpha$, by setting the derivatives to zero, the parameter ${\lambda}_i^{(l)}$
can be written as
\begin{align}
\lambda_i=\frac{1}{M}\Big\{\sum_{m=2}^M\Big[\!
-\!\frac{2\alpha}{1-\!\!\alpha^2}\Re\big\{\big[\boldsymbol\Pi_{m-1,m}^{(l)}
\big]_{i,i}\big\}\!+\!\frac{\alpha^2}{1-\alpha^2}
\big[\boldsymbol\Theta_{m-1}^{(l)}\big]_{i,i}\!+\!\frac{1}{1-\alpha^2}\big[\!\boldsymbol\Theta_{m}^{(l)}\big]_{i,i}
\Big]\!\!+\!\!\big[\boldsymbol\Theta_{1}^{(l)}\big]_{i,i}\!\Big\}.\label{eq:lambda_value}
\end{align}
On the other hand, with given $\lambda_i$, $\alpha$ can be achieved
through solving the following third-order equation~as
\begin{align}
&(M-1)N\alpha^3-\sum_{m=2}^M\Big(\tr\Big(\Re\Big\{{\boldsymbol\Lambda}^{-1}\boldsymbol\Pi_{m-1,m}^{(l)}\Big\}\Big)\Big)
\alpha^2
-\sum_{m=2}^M\Big(\tr\Big(\Re\Big\{{\boldsymbol\Lambda}^{-1}\boldsymbol\Pi_{m-1,m}^{(l)}\Big\}\Big)\Big)\notag\\
&+\left(\sum_{m=2}^M\Big(\tr\Big({\boldsymbol\Lambda}^{-1}\boldsymbol\Theta_{m-1}^{(l)}\Big)\Big)+
\sum_{m=2}^M\Big(\tr\Big({\boldsymbol\Lambda}^{-1}\boldsymbol\Theta_{m}^{(l)}\Big)\Big)-(M-1)N\right)\alpha=0.
\label{eq:alpha_value}
\end{align}
With (\ref{eq:lambda_value}) and (\ref{eq:alpha_value}), we can utilize
the fixed-point theorem to obtain $\widehat{\lambda}_i^{(l)}$ and $\widehat{\alpha}^{(l)}$.
Notice that the term $\boldsymbol\Pi_{m-1,m}^{(l)}$ can be written as
\begin{align}
\boldsymbol\Pi_{m-1,m}^{(l)}
=&\widehat{\tilde {\mathbf h}}_{m-1}\widehat{\tilde {\mathbf h}}_{m}^H
+\hat{\alpha}^{(l-1)}
\left(\boldsymbol\Theta_{m-1}^{(l)}-
         \widehat{\mathbf{\tilde h}}_{m-1}^{(l)}\left(\widehat{\mathbf{\tilde h}}_{m-1}^{(l)}\right)^H\right).
\end{align}
\begin{figure}[!t]
	\centering
	\includegraphics[width=3.5in]{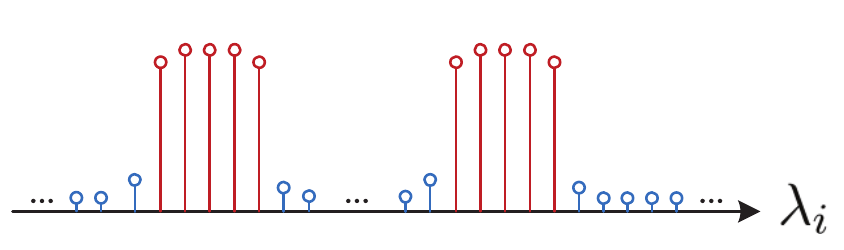}
	\caption{The spatial indices can be divided into two groups according to  $\lambda_i$.}
	\label{fig:k-means}
\end{figure}


\begin{algorithm}[t]
\setstretch{0.9}
	\caption{Obtaining   the non-zero  supporting vector from $\boldsymbol{\lambda}$ through  k-Means algorithm } 
	\begin{algorithmic}[1]\label{alg:k-means}
		\REQUIRE: $\boldsymbol{\lambda}$.
		\ENSURE: $\mathcal O$
		\STATE $\lambda_{c1} =\max ( \boldsymbol{\lambda}) $, $\lambda_{c2} =  \min(\boldsymbol{\lambda}_k )$,
		\REPEAT
			\STATE $ i=1 $ , $\mathcal O = \emptyset$ , $\mathcal O' = \emptyset$
			\REPEAT
			\IF {$|\lambda_{i}-\lambda_{c1} |> |\lambda_{i}-\lambda_{c2} |$}
			\STATE   $\mathcal O = \mathcal O \cup \{i\}$
			\ELSE
			\STATE  $\mathcal O' = \mathcal O' \cup \{i\}$
			\ENDIF
			\STATE $\lambda_{c1} = \frac{1}{|\mathcal O|} \sum_{j\in \mathcal O} \lambda_{j}$, $~\lambda_{c2} = \frac{1}{|\mathcal O|} \sum_{j\in \mathcal O'} \lambda_{j}$
			\UNTIL $i =N$
		\UNTIL $\mathcal O$ does not change.
	\end{algorithmic}
\end{algorithm}
\subsection{Obtain the Non-zero   Supporting Vector   from $\boldsymbol{\lambda}_k$}

As  mentioned in  the Section II, the virtual channel $\mathbf {h}_{m}$ is sparse. Let us use the set $\mathcal O$ to collect the indices of
the non-zero   elements of  $\mathbf {h}_{m}$ and refer to it as non-zero  supporting vector,   which is vital for the virtual channel tracking.
As shown in \figurename { \ref{fig:k-means}}, the spatial indices can be clearly classified into two groups according to the value of  $\lambda_i$.  The two groups correspond to  indices of zero elements  and the  none-zero elements, respectively. Based on  these   observations,  we resort to the  k-Means algorithm to efficiently extract spatial signature   from $\boldsymbol{\lambda}$. The detailed steps are shown in \textbf{Algorithm} \ref{alg:k-means}.

\section{Virtual Channel Tracking And Model Mismatch Detection}


\begin{algorithm}[!htp]
\setstretch{1.60}
    \caption{Channel tracking through GAMP} 
    \begin{algorithmic}[1]\label{alg:channel tracking}
    \REQUIRE: Learned parameters $\hat\alpha$, $\hat{\boldsymbol\lambda}$ and the Set $\mathcal Q$. Training matrix $\textbf{D}$, and observation vectors $\mathbf{y}_1^t$, $\mathbf y_2^t$, \ldots. The scalar estimation functions $g_{s}$ and $g_{\tilde h}$,and damping constants $\theta_{s}$,$\theta_{w}\in(0,1]$.
    \STATE  \textbf{Initialization:} $ m = 1$, $\boldsymbol{\mu}^t_{f_{1}^B\to  w_{1}}=0$, $\boldsymbol{\nu}^t_{f_{1}^B\to  w_{1}}=[\hat{\boldsymbol{\lambda}}]_{\mathcal{O}}$.
        \REPEAT
        \STATE $k=1$
           \REPEAT
           \STATE Implement the forward message passing into the $m$th time block, i=0.
        \REPEAT
        \STATE $i=i+1$
            \STATE  $\mu^t_{f_{m,i}^B\to  w_{m,i}}=
                     \hat\alpha\left(\frac{\bar\mu^t_{f_{m-1,:}^A\to w_{m-1,i}}}{\bar\nu^t_{f_{m-1,:}^A\to  w_{m-1,i}}}+\frac{\mu^t_{f_{m-1,i}^B\to  w_{m-1,i}}}{\nu^t_{f_{m-1,i}^B\to  w_{m-1,i}}}\right)\frac{\bar\nu^t_{f_{m-1,:}^A\to  w_{m-1,i}}\nu^t_{f_{m-1,i}^B\to  w_{m-1,i}}}{\bar\nu^t_{f_{m-1,i}^A\to  w_{m-1,i}}+\nu^t_{f_{m-1,i}^B\to  w_{m-1,i}}}$.
            \STATE $\nu^t_{f_{m,i}^B\to  w_{m,i}}=\left[\hat\alpha\right]^2\frac{\bar\nu^t_{f_{m-1,:}^A\to  w_{m-1,i}}\nu^t_{f_{m-1,i}^B\to  w_{m-1,i}}}{\bar\nu^t_{f_{m-1,:}^A\to w_{m-1,i}}+\nu^t_{f_{m-1,i}^B\to  w_{m-1,i}}}+\left(1-\left[\hat\alpha\right]^2\right)\left[[\hat{\boldsymbol{\lambda}}]_{\mathcal{O}}\right]_i.$¡£
        \UNTIL $i=|\mathcal O|$
        \STATE Implement the message exchanging within the $m$th time block.
        \STATE \textbf{GAMP}:
            \STATE $\mathbf{A}=\mathbf{D}^H.\mathbf{D}^H$(component-wise magnitude squared)
         \STATE 1./$\boldsymbol{\nu}_{p,m}^{t,(k)}=\mathbf{A}\boldsymbol{\nu}_{{w},m}^{t,(k)}$,\kern 20pt $\mathbf{p}_m^{t,(k)}=\mathbf{s}_m^{t,(k-1)}+\boldsymbol{\nu}_{p,m}^{t,(k)}.\mathbf{D}^H\mathbf{w}_m^{t,(k)}$.
         \STATE $\boldsymbol{\nu}_{s,m}^{t,(k)}=\boldsymbol{\nu}_{p,m}^{t,(k)}.g_{s}^{\prime}\left(\mathbf{p}_m^{{t,(k)}},\boldsymbol\nu_{p,m}^{t,(k)}\right)$,
         \kern 10pt $\mathbf s_m^{t,(k)}=(1-\theta_{s})\mathbf{s}_m^{t,(k-1)}+\theta_{s}g_{s}\left(\mathbf{p}_m^{t,(k)},\boldsymbol{\nu}_{p,m}^{t,(k)}\right)$.
         \STATE 1./${\boldsymbol\nu}_{r,m}^{t,(k)}=\mathbf{A}^{T}\boldsymbol{\nu}_{s,m}^{t,(k)}$,\kern 20pt $\mathbf{r}_m^{t,(k)}=\mathbf{w}_m^{t,(k)}-\boldsymbol{\nu}_{r,m}^{t,(k)}.\mathbf{D}\mathbf{s}_m^{t,(k)}$.
         \STATE $\boldsymbol{\tau}_{{w},m}^{t,(k+1)}=\boldsymbol{\nu}_{r,m}^{t,(k)}.g_{{w},m}^{\prime}(\mathbf{r}_m^{t,(k)},\boldsymbol{\nu}_{r,m}^{t,(k)})$,\kern 10pt $\mathbf{w}_m^{t,(k+1)}=(1-\theta_{w})\mathbf{w}_m^{{t,(k)}}+\theta_{w}g_{{w},m}(\mathbf{r}_m^{t,(k)},\boldsymbol{\nu}_{r,m}^{t,(k)})$.
         \STATE \textbf{GAMP END}.
        \STATE for $i=1$ to $i=|\mathcal O|$, $\bar\mu^t_{ f^A_{m,:}\to w_{m,i}}=[\mathbf r_{m}^{t,(k)}]_{i}$, $\bar\nu^t_{ f^A_{m,:}\to w_{m,i}}=[\boldsymbol\nu_{r,m}^{t,(k)}]_{i}$.
        \STATE $k=k+1$
        \UNTIL $k=K_{max}$
        \STATE $m=m+1$
        \UNTIL \text{The detection of the model mismatch.}
        \STATE Output the tracking results of the $m$th block, i.e.,
         $\hat{\textbf{w}}_{m}=\textbf{w}_m^{t,(K_{max}+1)},\boldsymbol {\Sigma}_{m}=\text{Diag}\left(\boldsymbol{\tau}_{w,m}^{t,(K_{max}+1)}\right)$.
    \end{algorithmic}
\end{algorithm}

\textcolor{black}{After the channel parameter leaning, each user can obtain
the information about $\boldsymbol\Xi=\{\alpha,\boldsymbol\Lambda\}$
and the corresponding supporting vector, denoted by $\mathcal O$.
With the uplink feedback link, the BS can collect the channel characteristics for
\cite{gao_E}, BS implements the user grouping according
to $\mathcal O$, and make sure that the supporting vectors for the users in the same group do not overlap.
Thus, the users in the same group can reuse the same training sequence. In
fact, after user grouping,}
%
	for the user with $\mathcal O$, only $|\mathcal O|$ orthogonal training sequences are required.
\textcolor{black}{With respect to different users' supporting index sets in one given group $\mathcal G_g$,
let us assume the biggest cardinality among all the related sets as $P_T$.}
So, we can build a   $P_T\times P_T$ matrix $\mathbf T_g $ with $\mathbf T_g  \mathbf T_g ^H = \frac{ \sigma_p^2 \mathbf I_{P_T}}{P_T}$
for this group,  and select $|\mathcal O|$ rows of   $\mathbf T_g $  to obtain $\mathbf D = [\mathbf T_g ]_{1:| {{\mathcal O} }|,:}.$
	Then, $\mathbf D$ is transmitted along the beam $[\mathbf F^H]_{:,\mathcal O}$,
and
	the received signal at  {the} user with $\mathcal O$ can be expressed as
	\begin{align}
	  { \mathbf y}_{m}^t= \mathcal{Q}\{\mathbf D^H [\tilde{\mathbf h}_{m}^t]_{\mathcal O}    + \mathbf n_{m}^t \},
	\end{align}
\textcolor{black}{where
the superscript denotes that the related variable belongs to the tracking phase;
$\mathbf y_{m}^t$, $\tilde{\mathbf h}_{m}^t$, and $\mathbf n_{m}^t$ separately have
the same meaning with respect to $\mathbf y_m$, $\tilde{\mathbf h}_m$, and $\mathbf n_m$ in (\ref{eq:dss_measure}}).
 Moreover, $\tilde{\mathbf h}_m$ and $\tilde{\mathbf h}_m^t$ (${\mathbf n}_m$ and ${\mathbf n}_m^t$)
 possess the same statistical characteristics.
 Then, we  {can obtain} the following state-space model  {as}
\begin{align}
[\tilde{\mathbf h}_{m}^t]_{\mathcal O} &= \hat{\alpha}  [\tilde{\mathbf h}_{m}^t]_{\mathcal O} +  \sqrt{1-\hat{\alpha}^2} \boldsymbol\upsilon_{m}^t,\label{eq:pf_state}\\
	  { \mathbf y}_{m}^t &= \mathcal{Q}\{\mathbf D^H [\tilde{\mathbf h}_{m}^t]_{\mathcal O}    + \mathbf n_{m}^t \},.\label{eq:pf_measure}
\end{align}
where the statistical characteristics of $\boldsymbol\upsilon_{m}^t$  is the same with
$\boldsymbol\upsilon_{m}$ in (\ref{eq:dss_state}).

Obviously, we can resort to the proper nonlinear filtering, for example,
the unscented Kalman filtering or the particle filtering,
to track the virtual channel $\mathbf w_m=[\mathbf {\tilde h}_m^t]_{\mathcal Q}$.
However, carefully analyzing the steps in \textbf{Algorithm} \ref{alg:GAMP_EM},
we can find that the message scheduling process is similar to the operations in
the Bayesian filtering and smoothing operations. Specially,
the forward message passing is equivalent to the filtering, while
the backward message passing is similar with the smoothing operation. Moreover, after
parameter learning, the state-space model in (\ref{eq:pf_state}) and (\ref{eq:pf_measure})
are low-dimensional without signal sparsity. Thus,
with the above observations, we will construct one GAMP-based virtual channel tracking scheme
\textcolor{black}{in \textbf{Algorithm} \ref{alg:channel tracking}}.
Notice that the notations $\mu^t_{f_{m,i}^B\to  w_{m,i}}$, $\nu^t_{f_{m,i}^B\to  w_{m,i}}$,
${\bar\mu}^t_{f_{m,i}^A\to  w_{m,i}}$, ${\bar\nu}^t_{f_{m,i}^A\to  w_{m,i}}$  in \textbf{Algorithm} \ref{alg:channel tracking} has
the similar meaning to $\mu_{f_{m,i}^B\to  {\tilde h}_{m,i}}$, $\nu_{f_{m,i}^B\to  {\tilde h}_{m,i}}$,
${\bar\mu}_{f_{m,i}^A\to  {\tilde h}_{m,i}}$, ${\bar\nu}_{f_{m,i}^A\to  {\tilde h}_{m,i}}$
in \textbf{Algorithm} \ref{alg:GAMP_EM}, respectively.
Correspondingly,
the explicit expressions for $g_{s}\left(\mathbf{p}^t,\boldsymbol{\nu}_{\mathbf p}^t\right)$,
$g_{w}(\mathbf{r}^t,\boldsymbol{\nu}_{\mathbf r}^t)$
can be inferred from
$g_{s}\left(\mathbf{p},\boldsymbol{\nu}_{\mathbf p}\right)$, $g_{\tilde h}(\mathbf{r},\boldsymbol{\nu}_{\mathbf r})$
 in Table. \ref{tab:g_s_h}
 through separately replacing ${\mu}_{f_{m,i}^B\to  w_{m,i}}$, ${\nu}_{f_{m,i}^B\to  w_{m,i}}$, ${\bar\mu}^t_{f_{m,i}^A\to  w_{m,i}}$, ${\bar\nu}^t_{f_{m,i}^A\to  w_{m,i}}$
 by $\mu^t_{f_{m,i}^B\to  w_{m,i}}$, $\nu^t_{f_{m,i}^B\to  w_{m,i}}$,
${\bar\mu}^t_{f_{m,i}^A\to  w_{m,i}}$, ${\bar\nu}^t_{f_{m,i}^A\to  w_{m,i}}$.

\begin{remark}
Due to the mobility of the users and change of environment, the learned parameters will
change in significant amounts. Thus, we have to start relearning process when the learned parameters
mismatch with the real scenario. Here, we can resort
to the Bayesian Cram\'er lower bound (BCRB) as the bench mark, which can be explained as follows.
After achieving the channel model parameters, we can construct
the tracking state space model as shown in (\ref{eq:pf_state}) and (\ref{eq:pf_measure}),
and derive the online BCRB.
At every time block, we   can get the virtual channel tracking MSE  from
the GAMP-based tracking scheme, which is denoted as $\boldsymbol\Sigma_m$ in
\textbf{Algorithm} \ref{alg:channel tracking}.
When   the tracking MSE is much higher than the corresponding BCRB,  it is considered that the  model parameters has changed and trigger the relearning process.
\end{remark}

\begin{figure}[!t]
	\centering
	\includegraphics[width=4.5in]{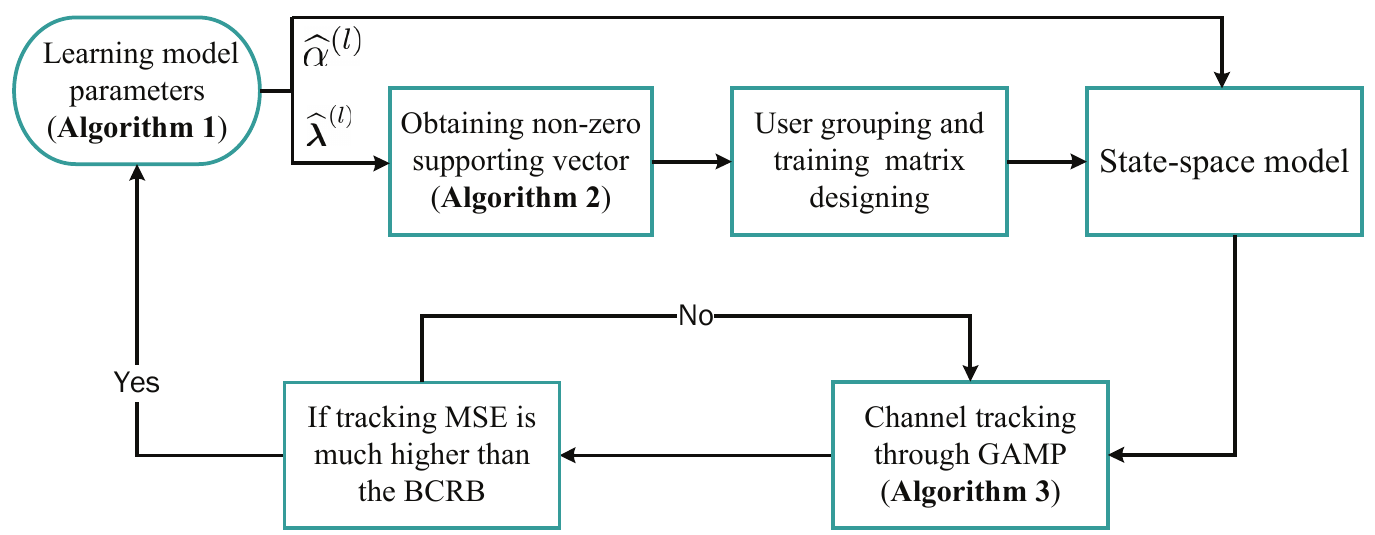}
	\caption{The   overall   block diagram of the proposed scheme.}
	\label{fig:diag_all}
\end{figure}

In order to describe the relationship among different parts
of the proposed scheme intuitively, the overall   block diagram of the proposed scheme are
illustrated in \figurename{ \ref{fig:diag_all}}.

\section{SIMULATION AND ANALYSIS}
In this section, we   evaluate the performance of our proposed scheme   through numerical simulation. The number of antennas at the BS is $N=128$. The BS antenna
spacing $d$ equals the half wavelength and the carrier frequency is
2GHz. The angular spread (AS) $\theta_k^{\text{max}}-\theta_k^{\text{min}}$ of the user is set as $4^\circ$  and the azimuth is randomly selected from $[-90^{\circ}, 90^{\circ}]$. We set the user's velocity as  100 km/h. The
	signal-to-noise ratio (SNR) is defined as SNR$= \sigma_p^2/\sigma_n^2$. $M=32$ channel coherent blocks are used to learn the   channel model parameters. The quantization step size is set according to \cite{7562390}.
Since the each of the users is  independent during the down-link channel estimation, we only consider one user here.

The   mean square error (MSE), which is formulated as follow,  is taken as performance metrics.
\begin{align}
\text{MSE}_{\mathbf x} =& \frac{1}{M} \sum_{m=1}^{M} \frac{\|\hat{\mathbf x}_m-\mathbf x_m\|^2}{\|\mathbf x_m\|^2}, {\mathbf x} = \alpha, \mathbf \Lambda,{\tilde{\mathbf h}}, {\mathbf w}.
\end{align}

\subsection { Model Parameters Learning}
\begin{figure}[!t]
	\centering
	\includegraphics[width=3.5in]{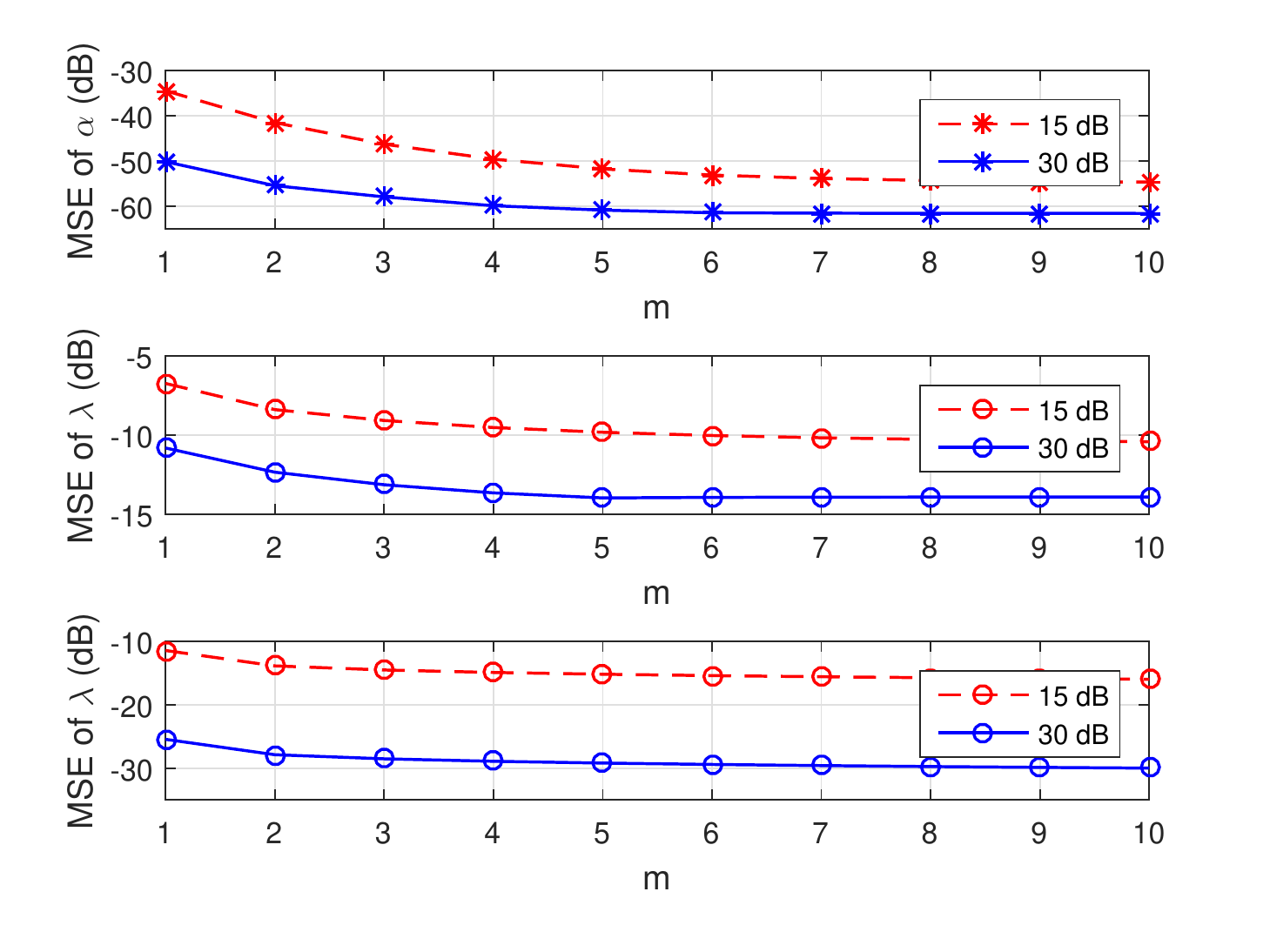}
	\caption{The convergence of the EM based model parameter learning algorithm (SNR=15dB and 30 dB).}
	\label{fig:EMparameter}
\end{figure}

We first investigate the convergence of the proposed GAMP-based model parameters learning scheme.
 The  MSE curves versus the number of    EM iteration
is shown in \figurename{ \ref{fig:EMparameter}}.  The  initial values are set as $\alpha_{k}=1$ and $\boldsymbol{\Lambda}_k = \mathbf I_{N}$ . It can be seen from \figurename{ \ref{fig:EMparameter}} that the EM algorithm  takes 8 and 6 iterations to arrive at  steady states for $\text{MSE}_{\alpha}$ and $\text{MSE}_{\boldsymbol{\Lambda}} $, respectively, under   SNR = 15dB. When the  SNR is 30dB, the convergence convergence speed is more faster.   We can also see that the $\text{MSE}_{\alpha}$  of the first iteration
	can be as low as -50dB. This is not strange because $\alpha_k$ ranges from   1 to 0.9899  for  a user with a velocity from $0$ km/h to $200$ km/h.  Therefore, the  initial value $\alpha_{k}=1$  is pretty  close to its true value.


	\begin{figure}[!t]
		\centering
		\includegraphics[width=3.5in]{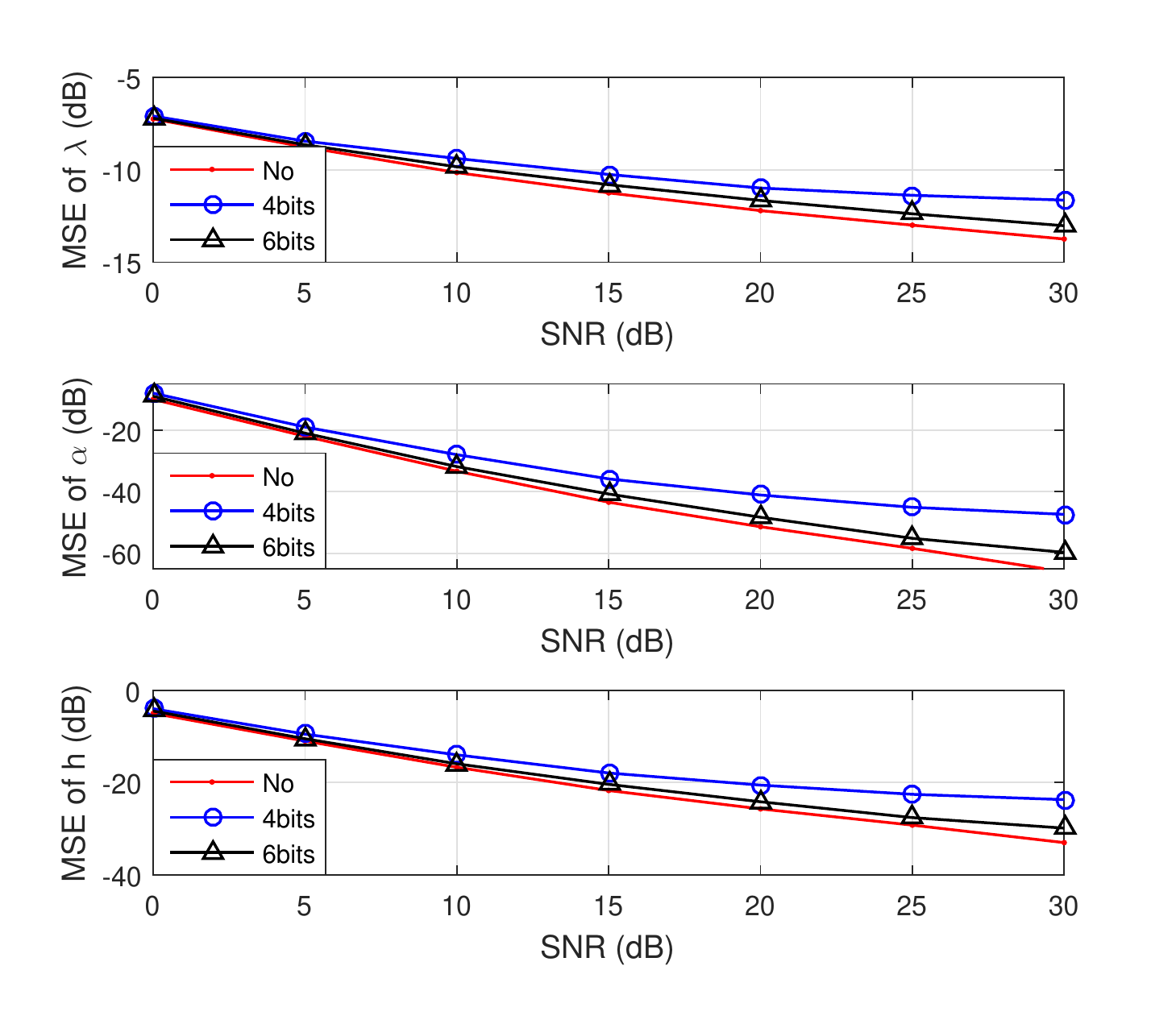}
		\caption{The  MSE performance of the model parameters learning versus  SNR. }
		\label{fig:model_SNR}
	\end{figure}
	
	\figurename{ \ref{fig:model_SNR}} presents the   MSE   of the model parameters learning as a function of the SNR.  10 iterations are used for the EM algorithm. 4-bit quantization, 6-bit quantization and no quantization cases are considered. We can see that even in the low SNR range, the   MSEs  of  $\alpha_k$  are very low while that of $\boldsymbol{\Lambda}_k$ is higher but still acceptable. 
	It is also shown in \figurename{ \ref{fig:model_SNR}}  that the MSE of  6-bit quantization is very close to that of no quantization, especially when  the SNR is low. Howerver, the MSE gap between the quantization and no quantization increase  with SNR. The reason behind this is that quantization will introduce some equivalent noise. Therefore, the higher the SNR, the greater the effect of quantization.

\begin{figure}[!t]
	\centering
	\includegraphics[width=3.7in]{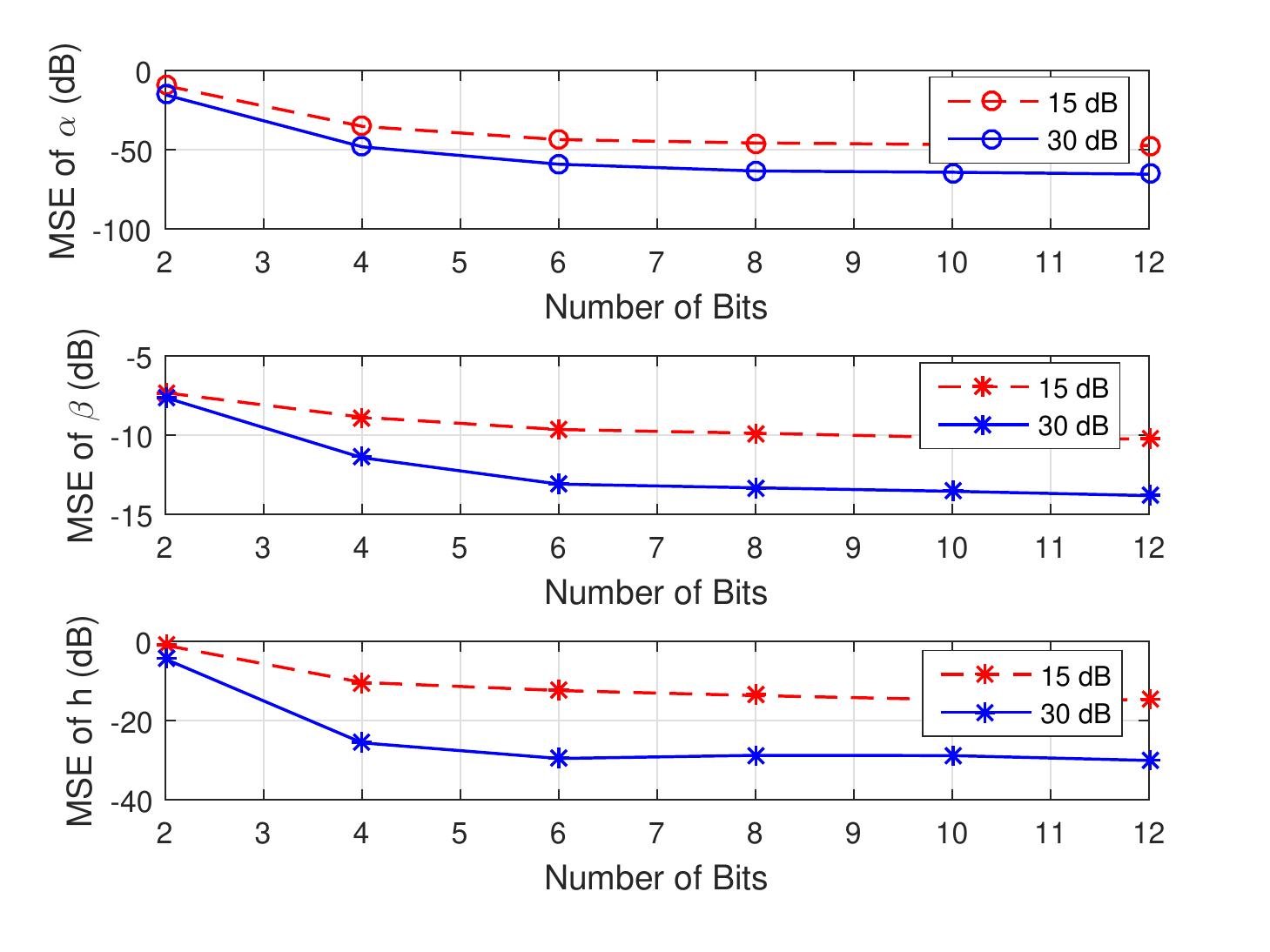}
	\caption{The MSE of the model parameters  versus the number of quantization bits (SNR=15 and 30 dB).}
	\label{fig:with_bit}
\end{figure}
To further  reveal the    quantization effects  on the, we present the MSE performance   of the model parametes learning versus the  the number of quantization bits  in \figurename{ \ref{fig:with_bit}}.  When the number of quantization bits is small, the equivalent noise introduced by  quantization is much higher than the noise. The  MSE can be evidently  reduced   by increasing number of quantization bits . However, when the number of quantization bits is large, the equivalent noise introduced by  quantization is much lower than the noise. The MSE is limited by the noise, especially when the SNR is low. Therefore, the MSE can not be deduced anymore by increasing number of quantization bits.
\subsection{ Low-dimensional Virtual Channel Tracking}

\begin{figure}[!t]
\centering
\subfigure[No quantization.]{
\begin{minipage}[t]{0.6\linewidth}
\centering
\includegraphics[width=3.6in]{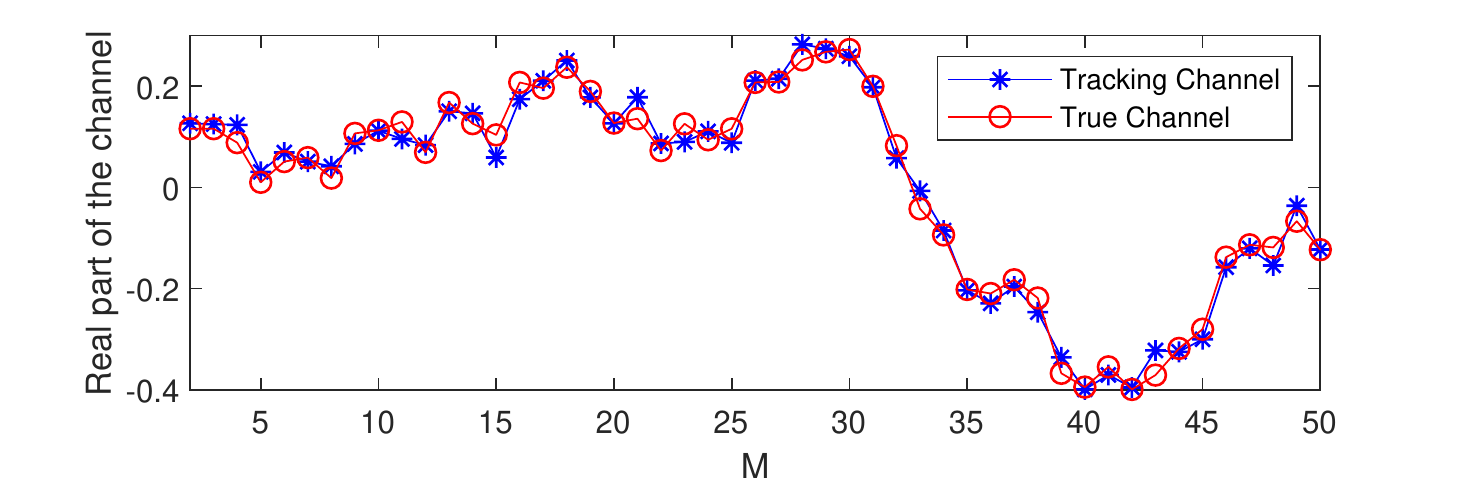}
\end{minipage}%
}
\subfigure[6-bits quantization.]{
\begin{minipage}[t]{0.6\linewidth}
\centering
\includegraphics[width=3.6in]{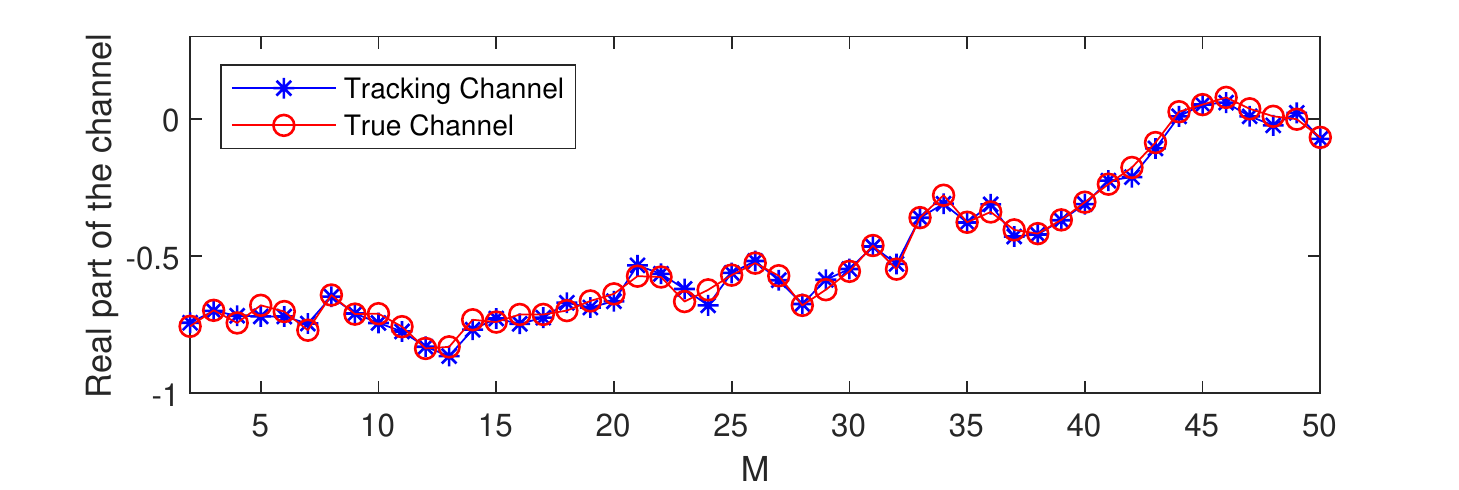}
\end{minipage}%
}%
\centering
\caption{Examples of channel tracking in the cases of non-quantization and 6-bits quantization.}
	\label{fig:tracing_example}
\end{figure}

Under the framework of  SBL, GAMP-based EM algorithm is used to learn the model parameters. After that, we will use the learned model parameters   to achieve low-dimensional virtual  channel tracking.
 Two examples of virtual channel tracking  for no quantization and 6-bit quantization are presented in \figurename{ \ref{fig:tracing_example}} presents. It can be seen that the curves  of the tracking result and true channel   are closely entangled, which explicitly shown that the performance of proposed low-dimensional virtual channel tracking  scheme is  satisfactory.

\begin{figure}[!t]
	\centering
	\includegraphics[width=3.6in]{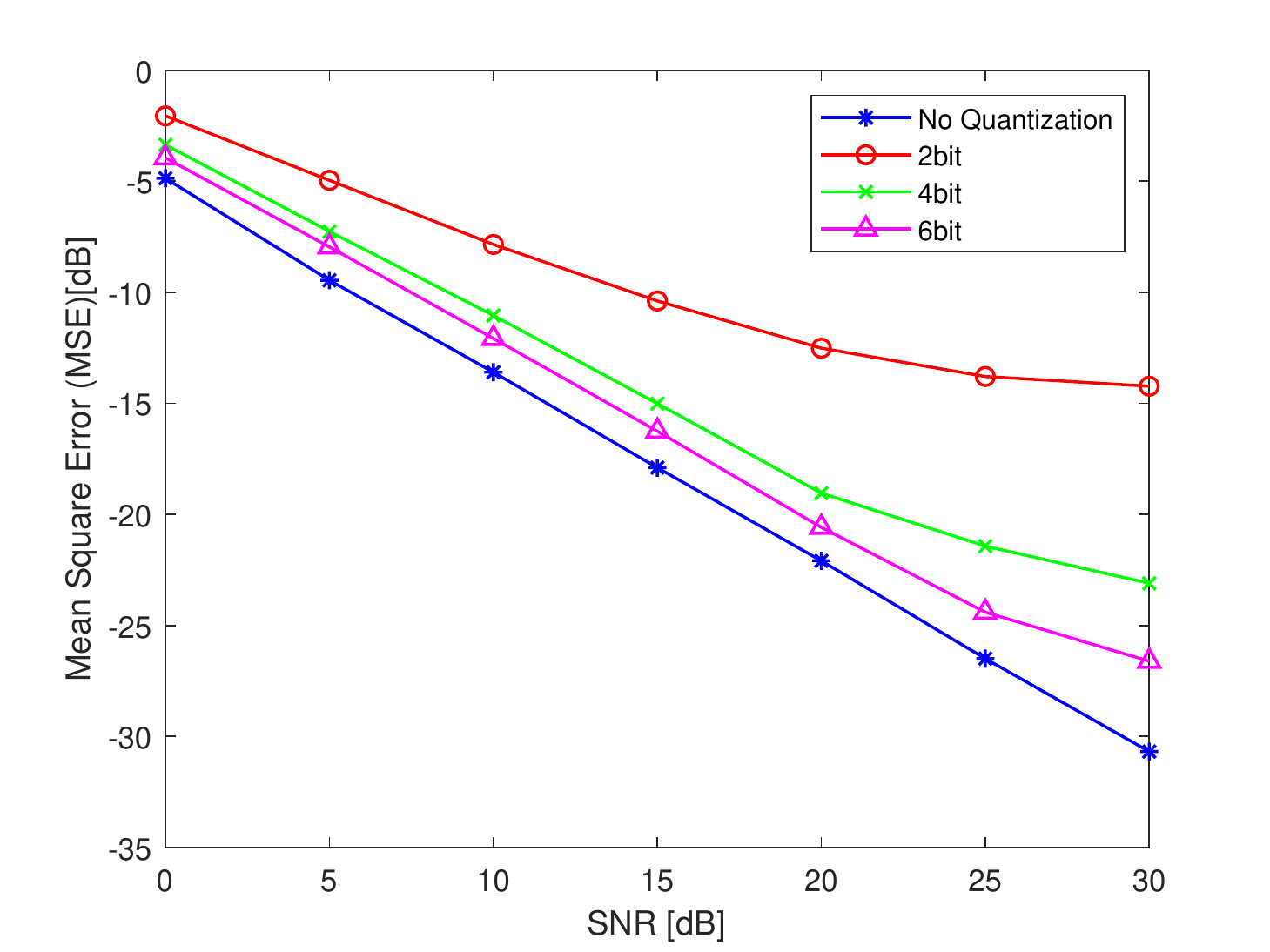}
	\caption{The performance of the virtual channel tracking versus SNR.}
	\label{fig:tracing_withSNR}
\end{figure}

Then, we present  the MSE of the proposed virtual
channel tracking method as a function of SNR in \figurename{ \ref{fig:tracing_withSNR}}. The MSEs  of 2-bit quantization, 4-bit quantization, 6-bit quantization and no  quantization are compared. The  performance trend over of virtual
channel tracking   likes that of model parameters learning. The MSEs of  6-bit quantization, 4-bit quantization and 2-bit quantization is  close to that of no quantization  when the SNR is low. Howerver, as the SNR increases,  the MSE gaps between different quantization become more and more larger.

\begin{figure}[!t]
	\centering
	\includegraphics[width=3.8in]{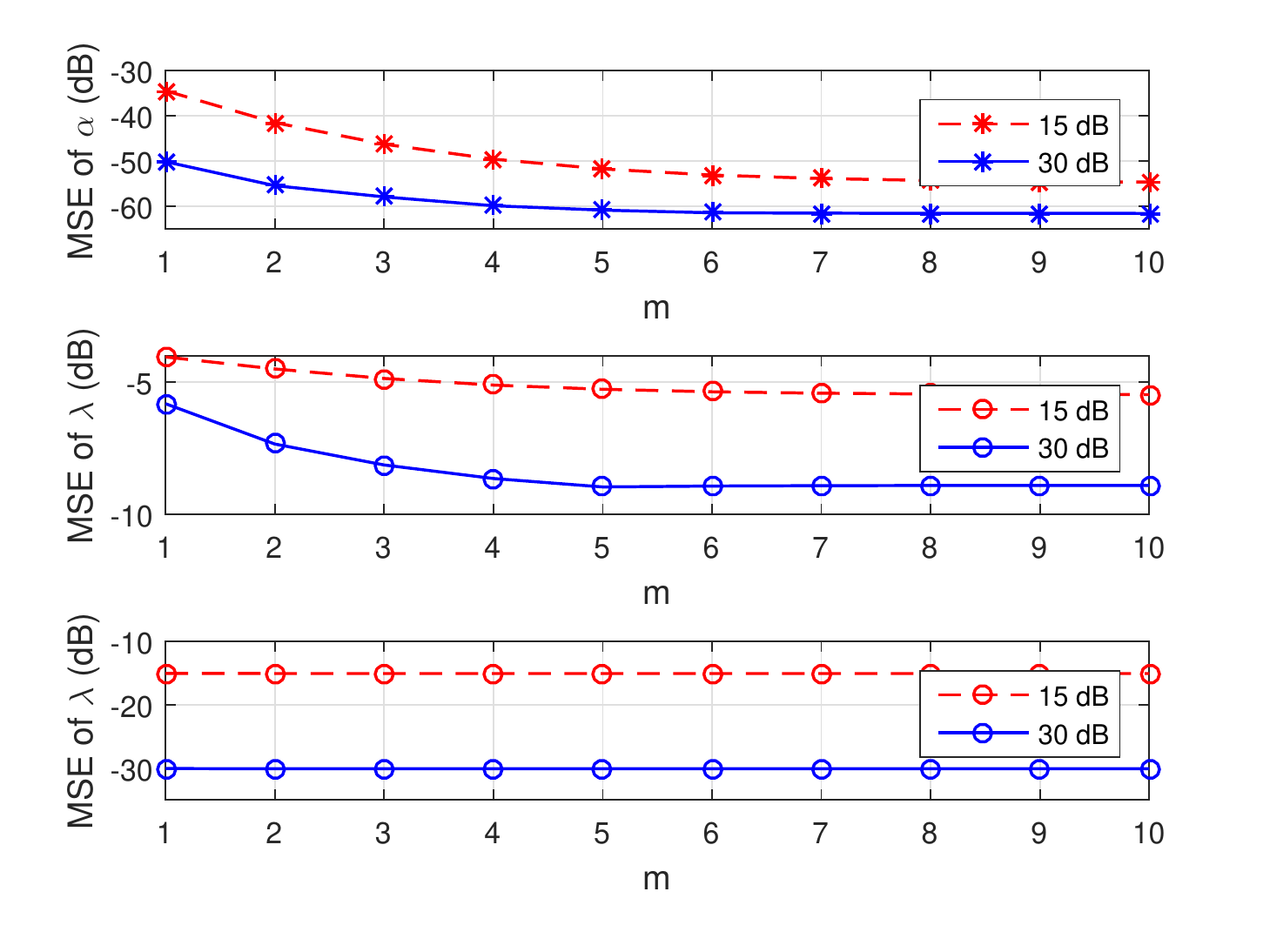}
	\caption{Comparison of   the virtual channel   tracking MSE and BCRB.}
	\label{fig:with_M}
\end{figure}

We display the   virtual
channel tracking  MSE curves versus number of tracking time blocks in \figurename{ \ref{fig:with_M}}. The online BCRB  is also presented as the bench mark.
It can be seen that both the BCRB and MSE    initially decrease and eventually converge as the time block increases. This is because the time correlation of the channel is used, which can improve the tracking accuracy in the latter time block. It can be also seen that the curves in the case of SNR = 30db  converge faster than in the case of SNR = 15dB. This is because the larger the SNR,     the more gain of using time-correlation, so the earlier the convergence.

\section{Conclusions}
In this paper, we proposed a Bayesian downlink channel estimation algorithm for the time-varying massive
MIMO networks. The effects of the quantization at the receiver are considered. Firstly, we developed an  EM  algorithm  based SBL framework to learn the model parameters of the sparse virtual channel. Specifically, the factor graph and the  GAMP  algorithms were used to compute the desired posterior statistics in the E-step.  Then, a reduced dimensional GAMP based scheme was proposed to track the virtual channel.
From the simulation results, the proposed model parameters learning algorithm  shows fast convergence speed. It  takes 8 and 6 iterations to arrive at  steady states for $\text{MSE}_{\alpha}$ and $\text{MSE}_{\boldsymbol{\Lambda}} $, respectively, under   SNR = 15dB. When the  SNR is higher, the convergence convergence speed is more faster.
The proposed virtual   channel tracking algorithm is able to  make the full use of the channel temporal correlation and enhance
the tracking accuracy.

\newpage
\section*{Appenxdix A\\
Calculation of $p\Big(y_{m,p}|z_{m,p};\widehat{\boldsymbol\Xi}^{(l-1)}\Big)$ and $p\Big(\tilde h_{m,i}|\tilde h_{m-1,i};\widehat{\boldsymbol\Xi}^{(l-1)}\Big) $}

Let us first consider the normal quantization in (\ref{eq:nor_quanti}).
From (\ref{eq:dss_measure}),
it can be readily checked that
$p\Big(q_{m,p}|z_{m,p};\widehat{\boldsymbol\Xi}^{(l-1)}\Big)=\mathcal{CN}\left(q_{m,p};{z_{m,p}},\sigma_n^2\right)$.
Hence, it can be checked that
\begin{align}
p\left(\Re\{q_{m,p}\}|\mathbf{\tilde h}_m;\alpha\right)=&\mathcal{N}
\left(\Re\{q_{m,p}\};\Re\{z_{m,p}\},\frac{\sigma_n^2}{2}\right),\\
p\left(\Im\{q_{m,p}\}|\mathbf{\tilde h}_m;\alpha\right)=&\mathcal{N}
\left(\Im\{q_{m,p}\};\Im\{z_{m,p}\},\frac{\sigma_n^2}{2}\right).
\end{align}
Before proceeding, let us define $\Phi(x)=\int^x_{-\infty}\frac{1}{\sqrt{2\pi}}e^{-\frac{u^2}{2}}du$.
With the property of Gaussian distributions, the conditional PDF
$p\Big(y_{m,p}|z_{m,p};\widehat{\boldsymbol\Xi}^{(l-1)}\Big)$ can be derived as
\begin{align}
p\Big(y_{m,p}|z_{m,p};\widehat{\boldsymbol\Xi}^{(l-1)}\Big)
=&F\left(\Re\{y_{m,p}\}\right) F\left(\Im\{y_{m,p}\}\right)
\end{align}
where $ F\left(\Re\{y_{m,p}\}\right)$ can be written as
\begin{small}
\begin{align}
F\left(\Re\{y_{m,p}\}\right)=\left\{
\begin{aligned}
&\Phi\left(\sqrt{2}\frac{(\Re\{y_{m,p}\}+\frac{1}{2})\triangle-\Re\{z_{m,p}\}}{\delta_n}\right)
\ ,\ \Re\{y_{m,p}\}<-\frac{2^{\kappa}}{2} \notag\\
&\Phi\left(\sqrt{2}\frac{(\Re\{y_{m,p}\}+\frac{1}{2})\triangle-\Re\{z_{m,p}\}}{\delta_n}\right)
-\notag\\ &\qquad \Phi\left(\sqrt{2}\frac{(\Re\{y_{m,p}\}-\frac{1}{2})\triangle-\Re\{z_{m,p}\}}{\delta_n}\right)
\ ,\ -\frac{2^{\kappa}}{2}\leqslant \Re\{y_{m,p}\}\leqslant \frac{2^{\kappa}}{2}-1 \notag\\
&1-\Phi\left(\sqrt{2}\frac{(\Re\{y_{m,p}\}-\frac{1}{2})\triangle-\Re\{z_{m,p}\}}{\delta_n}\right)
\ ,\ \Re\{y_{m,p}\}>\frac{2^{\kappa}}{2}-1
\end{aligned}
\right.,\label{eq:F_re}
\end{align}
\end{small}
and $ F\left(\Im\{y_{m,p}\}\right)$
can be achieved through  replacing $\Re\{y_{m,p}\}$ and
$\Re\{z_{m,p}\}$ in (\ref{eq:F_re}) with
$\Im\{y_{m,p}\}$ and
$\Im\{z_{m,p}\}$, respectively.

For the PDQ in (\ref{eq:pseu_quanti}), it can be derived that
$
y_{m,p}=(1-\rho)z_{m,p}+(1-\rho)n_{m,p}+n_{q,m,p}.
$
Thus, $p\Big(y_{m,p}|z_{m,p};\widehat{\boldsymbol\Xi}^{(l-1)}\Big)$ can be denoted as
\begin{align}
p\Big(y_{m,p}|z_{m,p};\widehat{\boldsymbol\Xi}^{(l-1)}\Big)=\mathcal{CN}(y_{m,p};(1-\rho)z_{m,p},(1-\rho)\sigma_n^2+\rho(1-\rho)).
\end{align}

From \eqref{eq:dss_state} it can be readily checked that
\begin{align}
 p\Big(\tilde h_{m,i}|\tilde h_{m-1,i};\widehat{\boldsymbol\Xi}^{(l-1)}\Big) \! =\!
\left\{\begin{aligned}
&\mathcal{CN}(\tilde h_{1,i};
0,\lambda_{i}^{(l-1)}), &~m=1,\\
& \mathcal{CN}\left(\tilde h_{m,i};
\hat\alpha^{(l-1)}\tilde h_{m-1,i},\Big(1-\Big[\hat\alpha^{(l-1)}\Big]^2\Big)\hat\lambda_{i}^{(l-1)}\right), &\text{otherwise}.
\end{aligned}
\right.
\end{align}

\newpage
\section*{Appendix B\\
The derivation  of $g_{s}\left(\mathbf{p},\boldsymbol{\nu}_{\mathbf p}\right)$
and $g_{\tilde h}(\mathbf{r},\boldsymbol{\nu}_{\mathbf r})$
}
In this appendix, we will derive the output scalar estimation
function $g_{s}\left(\mathbf{p},\boldsymbol{\nu}_{\mathbf p}\right)$
and that for the input  $g_{\tilde h}(\mathbf{r},\boldsymbol{\nu}_{\mathbf r})$,
where the prior distribution ( \ref{eq:GAMP_prior}) and the output function
 (\ref{eq:GAMP_mea}) will be utilized.

 With respect to the input function $g_{\tilde h}(\mathbf{r},\boldsymbol{\nu}_{\mathbf r})$,
 it can be verified that it is not dependent on the quantization scenario.
 With (\ref{eq:GAMP_prior}) and (\ref{eq:g_h_def}) , it can be derived
 \begin{small}
\begin{align}
&\int {\tilde h}_{m,i}p({\tilde h}_{m,i})\mathcal{CN}({\tilde h}_{m,i};r_i,\nu_{{r},i})
d{\tilde h}_{m,i}\notag\\
&\!=\!\!\frac{\mu_{\{f^B_{m+1,i}, f^B_{m,i}\}\to\tilde h_{m,i}}\nu_{{r},i}\!+\!r_i\nu_{\{f^B_{m+1,i}, f^B_{m,i}\}\to\tilde h_{m,i}}}{\nu_{\{f^B_{m+1,i}, f^B_{m,i}\}\to\tilde h_{m,i}}+\nu_{{r},i}}
\mathcal{CN}\left(\mu_{\{f^B_{m+1,i}, f^B_{m,i}\}\to\tilde h_{m,i}};r_i,
\nu_{\{f^B_{m+1,i}, f^B_{m,i}\}\to\tilde h_{m,i}}+\nu_{{r},i}\right),
\label{eq:g_h_deriv_one}\\
&\int\!\! p({\tilde h}_{m,i})\mathcal{CN}({\tilde h}_{m,i};r_i,\nu_{{r},i})d{\tilde h}_{m,i}
=\mathcal{CN}\left(\mu_{\{f^B_{m+1,i}, f^B_{m,i}\}\to\tilde h_{m,i}};r_i,\nu_{\{f^B_{m+1,i}, f^B_{m,i}\}\to\tilde h_{m,i}}+\nu_{{r},i}\right),\label{eq:g_h_deriv_two}\
\end{align}
\end{small}where the property equation
$\mathcal{CN}(x;\mu_1,\nu_1)\mathcal{CN}(x;\mu_2,\nu_2)=
\mathcal{CN}(x;\frac{\mu_1/\nu_1+\mu_2/\nu_2}{1/\nu_1+1/\nu_2},
\frac{1}{1/\nu_1+1/\nu_2})
\mathcal{CN}(0;\mu_1-\mu_2,\nu_1+\nu_2)$
are  utilized in the above derivations.
Hence,   the $i$th element of $g_{\tilde h}(\mathbf{r},\boldsymbol{\nu}_{\mathbf r})$ of
$g_{\tilde h}^{\prime}(\mathbf{r},\boldsymbol{\nu}_{\mathbf r})$  can be separately written as
\begin{align}
\left[g_{{\tilde h}}\left(\mathbf{r},\boldsymbol{\nu}_{\mathbf r}\right)\right]_i=&
\frac{\mu_{\{f^B_{m+1,i}, f^B_{m,i}\}\to\tilde h_{m,i}}\nu_{{r},i}+r_i\nu_{\{f^B_{m+1,i}, f^B_{m,i}\}\to\tilde h_{m,i}}}{\nu_{\{f^B_{m+1,i}, f^B_{m,i}\}\to\tilde h_{m,i}}+\nu_{{r},i}}, \label{eq: g_h_exp}\\
\left[g_{{\tilde h}}^\prime\left(\mathbf{r},\boldsymbol{\nu}_{\mathbf r}\right)\right]_i=&
=\frac{\nu_{\{f^B_{m+1,i}, f^B_{m,i}\}\to\tilde h_{m,i}}}{\nu_{\{f^B_{m+1,i}, f^B_{m,i}\}\to\tilde h_{m,i}}+\nu_{{r},i}}.
\end{align}
The term  $g_{s}\left(\mathbf{p},\boldsymbol{\nu}_{\mathbf p}\right)$
will be further examined.  First, we will not consider the quantization effect,
and can obtain $p(y_{m,p}|z_{m,p})=\mathcal{CN}(y_{m,p};z_{m,p},\sigma_n^2)$.
Then, from (\ref{eq:g_s_def}), we have
\begin{align}
\int \!\!d{z}_{m,n}z_{m,n}p({y_{m,n}|z_{m,n}})\mathcal{CN}\left({z}_{m,n};\frac{p_n}{\nu_{ p,n}},\frac{1}{\nu_{p,n}}\right)
=&\frac{\frac{y_{m,n}}{\sigma_n^2}+p_n}{\frac{1}{\sigma_n^2}+{\nu_{p,n}}}
 \mathcal{CN}\left(y_{m,n};\frac{p_n}{\nu_{p,n}},\sigma_n^2+\frac{1}{\nu_{p,n}}\right),\\
\int d{z}_{m,n}p({y_{m,n}|z_{m,n}})\mathcal{CN}\left({z}_{m,n};\frac{p_n}{\nu_{ p,n}},\frac{1}{\nu_{p,n}}\right)=&\mathcal{CN}\left(y_{m,n};\frac{p_n}{\nu_{ p,n}},\sigma_n^2+\frac{1}{\nu_{p,n}}\right),
\end{align}
where the calculation techniques in (\ref{eq:g_h_deriv_one}) and ((\ref{eq:g_h_deriv_two}) )
are utilized in the above derivations. Then, the $n$th entry of $g_{s}\left(\mathbf{p},\boldsymbol{\nu}_{\mathbf p}\right)$
and $g_{s}^\prime\left(\mathbf{p},\boldsymbol{\nu}_{\mathbf p}\right)$ can be denoted as
\begin{align}
\left[g_{\mathbf s}\left(\mathbf{p},\boldsymbol{\nu}_{\mathbf p}\right)\right]_n
=\frac{p_n-\nu_{p,n}{y_{m,n}}}{1+\nu_{p,n}{\sigma_n^2}},\kern 20pt
\left[g_{\mathbf s}^\prime\left(\mathbf{p},\boldsymbol{\nu}_{\mathbf p}\right)\right]_n=\frac{1}{1+\nu_{p,n}{\sigma_n^2}}.
\end{align}
Following the similar case, we can derive
$g_{s}\left(\mathbf{p},\boldsymbol{\nu}_{\mathbf p}\right)$
and $g_{s}^\prime\left(\mathbf{p},\boldsymbol{\nu}_{\mathbf p}\right)$ under
the PDQ scenario as
\begin{align}
\left[g_{\mathbf s}\left(\mathbf{p},\boldsymbol{\nu}_{\mathbf p}\right)\right]_n
=&\frac{(1-\rho)p_n-\nu_{p,n}y_{m,n}}{(1-\rho)+\nu_{p,n}[(1-\rho)\sigma_n^2+\rho]},
\end{align}
\begin{align}
\left[g_{\mathbf s}^\prime\left(\mathbf{p},\boldsymbol{\nu}_{\mathbf p}\right)\right]_n=&\frac{1-\rho}{(1-\rho)+\nu_{p,n}[(1-\rho)\sigma_n^2+\rho]}.
\end{align}

With respect to the normal quantization case,  we can obtain
\begin{align}
&\int p({y_{m,n}|z_{m,n}})\mathcal{CN}\left({z}_{m,n};\frac{p_n}{\nu_{p,n}},\frac{1}{\nu_{p,n}}\right)
d{z}_{m,n}\notag\\
&=\int_{\epsilon_{\Re\{y_{m,n}\}}^L}^{\epsilon_{\Re\{y_{m,n}\}}^U}
\int_{\epsilon_{\Im\{y_{m,n}\}}^L}^{\epsilon_{\Im\{y_{m,n}\}}^U} \mathcal{CN}\left({q}_{m,n};\frac{p_n}{\nu_{p,n}},\sigma_n^2+\frac{1}{\nu_{p,n}}\right)d{q}_{m,n} \notag\\ &=\left({\Phi(\zeta_{m,n})-\Phi(\eta_{m,n})}\right)\left({\Phi(\bar\zeta_{m,n})-\Phi(\bar\eta_{m,n})}\right),
\label{eq:gs_normal_cal_one}
\end{align}
where the terms $\zeta_{m,n}=\frac{\epsilon_{\Re\{y_{m,n}\}}^U-\frac{\Re\{p_n\}}{\nu_{p,n}}}{\sqrt{\frac{1}{2}\left(\sigma_n^2+\frac{1}{\nu_{p,n}}\right)}}$,
$\eta_{m,n}=\frac{\epsilon_{\Re\{y_{m,n}\}}^L-\frac{\Re\{p_n\}}{\nu_{p,n}}}{\sqrt{\frac{1}{2}\left(\sigma_n^2+\frac{1}{\nu_{p,n}}\right)}}$,
$\bar\zeta_{m,n}=\frac{\epsilon_{\Im\{y_{m,n}\}}^U-\frac{\Im\{p_n\}}{\nu_{p,n}}}{\sqrt{\frac{1}{2}\left(\sigma_n^2+\frac{1}{\nu_{p,n}}\right)}}$,
$\bar\eta_{m,n}=\frac{\epsilon_{\Im\{y_{m,n}\}}^L-\frac{\Im\{p_n\}}{\nu_{p,n}}}{\sqrt{\frac{1}{2}\left(\sigma_n^2+\frac{1}{\nu_{p,n}}\right)}}$
are defined in the above equation.

Then, with respect to $\int z_{m,n}p({y_{m,n}|z_{m,n}})\mathcal{CN}\left({z}_{m,n};\frac{p_n}{\nu_{p,n}},\frac{1}{\nu_{p,n}}\right)
d{z}_{m,n}$, we can obtain
\begin{align}
&\int z_{m,n}p({y_{m,n}|z_{m,n}})\mathcal{CN}\left({z}_{m,n};\frac{p_n}{\nu_{p,n}},\frac{1}{\nu_{p,n}}\right)
d{z}_{m,n}\notag\\
=&\int_{\epsilon_{\Re\{y_{m,n}\}}^L}^{\epsilon_{\Re\{y_{m,n}\}}^U}
\int_{\epsilon_{\Im\{y_{m,n}\}}^L}^{\epsilon_{\Im\{y_{m,n}\}}^U}\frac{{q}_{m,n}+\sigma_n^2 p_n}{1+\sigma_n^2 \nu_{p,n}}
\mathcal{CN}\left({q}_{m,n};\frac{p_n}{\nu_{p,n}},\sigma_n^2+\frac{1}{\nu_{p,n}}\right) d{q}_{m,n}\notag\\
=&\frac{\left({\Phi(\zeta_{m,n})-\Phi(\eta_{m,n})}\right)\left({\Phi(\bar\zeta_{m,n})-\Phi(\bar\eta_{m,n})}\right)}{1+\sigma_n^2\nu_{p,n}}\left\{\left[\mathbb E_{\mathcal{TN}}\{\Re\{q_{m,n}\}\}+\jmath
\mathbb E_{\mathcal{TN}}\{\Im\{q_{m,n}\}\}\right]+\sigma_n^2p_n\right\},
\label{eq:gs_normal_cal_two}
\end{align}
where both $\Re\{q_{m,n}\}$
and $\Im\{q_{m,n}\}$ are truncated normal distributed \cite{williams2006gaussian},
and their  PDFs are
\begin{align}
p(\Re\{q_{m,n}\})=&\mathcal{TN}\left(\Re\{q_{m,n}\};\frac{\Re\{p_n\}}{\nu_{p,n}},\frac{1}{2}\left(\sigma_n^2+\frac{1}{\nu_{p,n}}\right),\epsilon_{\Re\{y_{m,n}\}}^L,\epsilon_{\Re\{y_{m,n}\}}^U\right),\\
p(\Im\{q_{m,n}\})=&\mathcal{TN}\left(\Im\{q_{m,n}\};\frac{\Im\{p_n\}}{\nu_{p,n}},\frac{1}{2}\left(\sigma_n^2+\frac{1}{\nu_{p,n}}\right),\epsilon_{\Im\{y_{m,n}\}}^L,\epsilon_{\Im\{y_{m,n}\}}^U\right).
\end{align}
Moreover, the notation $\mathcal{TN}(x;\mu,\nu^2,a,b)$ denotes the real random variable $x$ is truncated normal distributed in the region $[a,b]$, where the non-truncated version of $x$ is normal distributed with mean $\mu$
and variance $\nu^2$.
Plugging (\ref{eq:gs_normal_cal_one}) and (\ref{eq:gs_normal_cal_two}) into
(\ref{eq:g_s_def}), we can obtain the $n$th element of $\left[g_{s}\left(\mathbf{p},\boldsymbol{\nu}_{p}\right)\right]_n$
under the normal quantization case as
\begin{align}
\left[g_{s}\left(\mathbf{p},\boldsymbol{\nu}_{p}\right)\right]_n=&p_n-
\frac{\nu_{p,n}}{1+\sigma_n^2\nu_{p,n}}\left\{\left[\mathbb E_{\mathcal{TN}}\{\Re\{q_{m,n}\}\}+\jmath
\mathbb E_{\mathcal{TN}}\{\Im\{q_{m,n}\}\}\right]+\sigma_n^2p_n\right\}.
\label{eq:gs_normal_exp_one}
\end{align}

With the theory of the truncated normal distribution, the following equation can be obtained
\begin{align}
\int_{a}^b x \mathcal{TN}(x;\mu,\nu^2,a,b) dx=\mu+\nu^2\frac{\mathcal N(a;\mu,\nu^2)-\mathcal N(b;\mu,\nu^2)}
{\Phi\left(\frac{b-\mu}{\nu}\right)-\Phi\left(\frac{a-\mu}{\nu}\right)}.
\end{align}
With this property, we can rewrite (\ref{eq:gs_normal_exp_one}) as
\begin{align}
\left[\!g_{s}\!\left(\mathbf{p},\!\boldsymbol{\nu}_{p}\right)\!\right]_n\!\!=&p_n\!\!-\!\!
\frac{\nu_{p,n}}{1\!+\!\sigma_n^2\nu_{p,n}}\!\left\{\!\left[\mathbb E_{\mathcal{TN}}\{\Re\{q_{m,n}\}\}\!+\!\jmath
\mathbb E_{\mathcal{TN}}\{\!\Im\{q_{m,n}\!\}\!\}\!\right]\!+\!\sigma_n^2p_n\right\}
\!=\!\frac{1}{2}\!\frac{\Delta^{\Re}_{m,n}}{\nabla^{\Re}_{m,n}}
\!\!+\!\!\frac{1}{2}\jmath\frac{\Delta^{\Im}_{m,n}}{\nabla^{\Im}_{m,n}},
\label{eq:gs_normal_exp_two}
\end{align}
where $\nabla^{\Re}_{m,n}={\Phi(\zeta_{m,n})-\Phi(\eta_{m,n})}$,
$\nabla^{\Im}_{m,n}={\Phi(\bar\zeta_{m,n})-\Phi(\bar\eta_{m,n})}$,
and $\Delta^{\Re}_{m,n}$, $\Delta^{\Im}_{m,n}$ can be separately written as
\begin{align}
\Delta^{\Re}_{m,n}=\!\mathcal{N}\left(\epsilon_{\Re\{y_{m,n}\}}^U;\frac{\Re\{p_n\}}{\nu_{p,n}},\frac{1}{2}\left(\sigma_n^2+\frac{1}{\nu_{p,n}}\right)\right)
-\!\mathcal{N}\left(\epsilon_{\Re\{y_{m,n}\}}^L;\frac{\Re\{p_n\}}{\nu_{p,n}},\frac{1}{2}\left(\sigma_n^2+\frac{1}{\nu_{p,n}}\right)\right),\\
\Delta^{\Im}_{m,n}=\!\mathcal{N}\left(\epsilon_{\Im\{y_{m,n}\}}^U;\frac{\Im\{p_n\}}{\nu_{p,n}},\frac{1}{2}\left(\sigma_n^2+\frac{1}{\nu_{p,n}}\right)\right)
-\!\mathcal{N}\left(\epsilon_{\Im\{y_{m,n}\}}^L;\frac{\Im\{p_n\}}{\nu_{p,n}},\frac{1}{2}\left(\sigma_n^2+\frac{1}{\nu_{p,n}}\right)\right).
\end{align}
Thus, the $n$th element of $g_{s}^\prime\left(\mathbf{p},\boldsymbol{\nu}_{\mathbf p}\right)$
under the normal quantization case can be derived as
\begin{align}
[g_{s}^\prime\left(\mathbf{p},\boldsymbol{\nu}_{\mathbf p}\right)]_n=
\frac{1}{4}\Bigg\{\frac{\frac{\partial\Delta^{\Re}_{m,n}}{\partial\Re\{p_n\}}
\nabla^{\Re}_{m,n}-\Delta^{\Re}_{m,n}
\frac{\partial\nabla^{\Re}_{m,n}}{\partial\Re\{p_n\}}}{(\nabla^{\Re}_{m,n})^2}
+\frac{\frac{\partial\Delta^{\Im}_{m,n}}{\partial\Im\{p_n\}}
\nabla^{\Im}_{m,n}-\Delta^{\Im}_{m,n}
\frac{\partial\nabla^{\Im}_{m,n}}{\partial\Im\{p_n\}}}{(\nabla^{\Im}_{m,n})^2}\Bigg\}.
\label{eq:gs_der_normal_one}
\end{align}
After some calculations, we have
\begin{small}
\begin{align}
&\frac{\partial\Delta^{\Re}_{m,n}}{\partial\Re\{p_n\}}
\!=\!\!\frac{1}{\nu_{p,n}\sqrt{\!\!\frac{1}{2}\!\!\left(\!\!\sigma_n^2\!+\!\!\frac{1}{\nu_{p,n}}\!\!\right)}}\! \Big(\!\!\underbrace{\mathcal N\!\!\left(\!\epsilon_{\Re\{y_{m,n}\}}^U;\!\frac{\Re\{p_n\}}{\nu_{p,\!n}},\!\frac{1}{2}\!\!\left(\!\sigma_n^2\!+\!\!\frac{1}{\nu_{p,n}}\!\!\right)\!\!\right)\zeta_{m,n}
\!\!-\!\!\mathcal N\!\left(\!\epsilon_{\Re\{y_{m,n}\}}^L;\frac{\Re\{\!p_n\}}{\nu_{p,n}},\frac{1}{2}\!\!\left(\!\sigma_n^2\!+\!\!\frac{1}{\nu_{p,n}}\!\!\right)\!\!\right)
\eta_{m,n}}_{\Xi^{\Re}_{m,n}}\Big),\notag \\
&\frac{\partial\Delta^{\Im}_{m,n}}{\partial\Im\{p_n\}}
\!=\!\!\frac{1}{\nu_{p,n}\sqrt{\frac{1}{2}\!\!\left(\!\sigma_n^2\!\!+\!\!\frac{1}{\nu_{p,\!n}}\!\!\right)}} \Big(\!\!\underbrace{\mathcal N\!\!\left(\!\!\epsilon_{\!\Im\{y_{m,n}\!\}}^U;\!\!\frac{\Im\{p_n\}}{\nu_{p,n}},\frac{1}{2}\!\!\left(\!\!\sigma_n^2\!+\!\frac{1}{\nu_{p,n}}\!\!\right)\!\right)
\bar\zeta_{m,n}\!-\!\mathcal N\!\!\left(\!\epsilon_{\Im\{y_{m,n}\}}^L;\frac{\Im\{p_n\}}{\nu_{p,n}},\frac{1}{2}\!\!\left(\!\!\sigma_n^2\!\!+\!\!\frac{1}{\nu_{p,n}}\!\right)\!\right)
\bar\eta_{m,n}}_{\Xi^{\Im}_{m,n}}\!\Big),\notag\\
&\frac{\partial\nabla^{\Re}_{m,n}}{\partial\Re\{p_n\}}
=-\frac{1}{\nu_{p,n}}\underbrace{\left(\mathcal N\left(\epsilon_{\Re\{y_{m,n}\}}^U;\frac{\Re\{p_n\}}{\nu_{p,n}},\frac{1}{2}\left(\sigma_n^2+\frac{1}{\nu_{p,n}}\right)\right)
-\mathcal N\left(\epsilon_{\Re\{y_{m,n}\}}^L;\frac{\Re\{p_n\}}{\nu_{p,n}},\frac{1}{2}\left(\sigma_n^2+\frac{1}{\nu_{p,n}}\right)\right)\right)}_{\Delta^{\Im}_{m,n}},\notag\\
&\frac{\partial\nabla^{\Im}_{m,n}}{\partial\Im\{p_n\}}
=-\frac{1}{\nu_{p,n}}\underbrace{\left(\mathcal N\left(\epsilon_{\Im\{y_{m,n}\}}^U;\frac{\Im\{p_n\}}{\nu_{p,n}},\frac{1}{2}\left(\sigma_n^2+\frac{1}{\nu_{p,n}}\right)\right)
-\mathcal N\left(\epsilon_{\Im\{y_{m,n}\}}^L;\frac{\Im\{p_n\}}{\nu_{p,n}},\frac{1}{2}\left(\sigma_n^2+\frac{1}{\nu_{p,n}}\right)\right)\right)}_{\Delta^{\Im}_{m,n}},\notag
\end{align}
\end{small}where the terms $\boldsymbol\Xi^{\Re}_{m,n}$ and
$\boldsymbol\Xi^{\Im}_{m,n}$ are defined in the above equations.
Substituting the above partial derivatives into
(\ref{eq:gs_der_normal_one}), we can achieve
\begin{align}
[g_{s}^\prime\left(\mathbf{p},\boldsymbol{\nu}_{\mathbf p}\right)]_n
=\frac{1}{4\nu_{p,n}\sqrt{\frac{1}{2}\left(\sigma_n^2+\frac{1}{\nu_{p,n}}\right)}}
\left\{\frac{\Xi^{\Re}_{m,n}}{\nabla^{\Re}_{m,n}}+\frac{\Xi^{\Im}_{m,n}}{\nabla^{\Im}_{m,n}} \right\}
+\frac{1}{\nu_{p,n}}|\left[g_{s}\left(\mathbf{p},\boldsymbol{\nu}_{p}\right)\right]_n|^2
\label{eq:gs_der_normal_two}
\end{align}

\balance
\bibliographystyle{IEEEtran}
\bibliography{./bibtex/IEEEabrv,./bibtex/ref}
\end{document}